\documentclass[12pt,draftclsnofoot,onecolumn,romanappendices]{IEEEtran}
 
\usepackage[T1]{fontenc}		
\usepackage[utf8]{inputenc}	

\usepackage{cite}
\usepackage{graphicx}
\usepackage{amsmath,amssymb}
\usepackage{amsfonts}
\usepackage{amsthm}

\usepackage{comment}
\usepackage{epstopdf}
\usepackage{color}

\usepackage{hyperref}
\usepackage{subcaption}
\usepackage{nicefrac}
\usepackage{placeins} 
\usepackage[ruled,vlined,linesnumbered]{algorithm2e}
\usepackage[dvipsnames]{xcolor}

\usepackage{microtype}

\newtheorem{remark}{Remark}

\newtheorem{lemma}{Lemma}
\newtheorem{prop}{Proposition}
\newtheorem{coro}{Corollary}

\newtheorem{example}{Example}

\usepackage{hyperref} 
\hypersetup{
	colorlinks = true,
	linkcolor  = red,
	citecolor  = green!80!black,
	urlcolor   = black
}

\usepackage{url}

	\newcommand{\dd}{\mathrm{d}}

	\newcommand{\E}{\mathbb{E}}
	\newcommand{\Zb}{\mathbb{Z}}

	\newcommand{\Cc}{\mathcal{C}}
	\newcommand{\Gc}{\mathcal{G}}

	\newcommand{\Nc}{\mathcal{N}}
	\newcommand{\Tc}{\mathcal{T}}
	\newcommand{\Xc}{\mathcal{X}}
	\newcommand{\Wc}{\mathcal{W}}
	\newcommand{\Yc}{\mathcal{Y}}
	\newcommand{\Zc}{\mathcal{Z}}
	
	\newcommand{\bb}{\mathbf{b}}
	
	\newcommand{\bv}{\mathbf{v}}
	
	\DeclareMathOperator{\Ei}{\mathrm{Ei}}
	\DeclareMathOperator{\Exp}{\mathrm{Exp}}
	
	\DeclareMathOperator{\SNR}{\mathrm{SNR}}
	
	\DeclareMathOperator*{\argmax}{arg\,max}

\newcommand{\bR}{\Bar{R}}
\newcommand{\emn}{{\rm (MN)}}
\newcommand{\eacc}{{\rm (ACC)}}

\newcommand{\rmn}{\Bar{R}^{\rm (MN)}}
\newcommand{\racc}{\Bar{R}^{\rm (ACC)}}

\definecolor{color1}{rgb}{0.00000,0.44700,0.74100}%
\colorlet{myred}{red!80!black}%
\colorlet{myblue}{color1!80!black}%

\begin{document}

\title{\LARGE{Wireless Coded Caching Can Overcome the Worst-User Bottleneck by Exploiting Finite File Sizes}}
\author{Hui Zhao, Antonio Bazco-Nogueras, and Petros Elia \vspace{-2cm}
	\thanks{The authors are with the Communication Systems Department, EURECOM, 06410 Sophia Antipolis, France (email: hui.zhao@eurecom.fr; antonio.bazco-nogueras@eurecom.fr; petros.elia@eurecom.fr). This work is supported by the European Research Council under the EU Horizon 2020 research and innovation program / ERC grant agreement no. 725929 (ERC project DUALITY). Part of this work has been accepted to the 2020 IEEE Information Theory Workshop\cite{Zhao2020_ITW}.}}

\maketitle

	\begin{abstract}
		We address the worst-user bottleneck of wireless coded caching, which is known to severely diminish cache-aided multicasting gains due to the fundamental worst-channel limitation of multicasting transmission. 
		We consider the quasi-static Rayleigh fading Broadcast Channel, for which we first show that the effective coded caching gain of the XOR-based standard coded-caching scheme completely vanishes in the low-SNR regime. 
		Then, we reveal that this collapse is not intrinsic to coded caching. We do so by presenting a novel scheme that can fully recover the coded caching gains by capitalizing on one aspect that has to date remained unexploited: the shared side information brought about by the effectively unavoidable file-size constraint. As a consequence, the worst-user effect is dramatically ameliorated, as it is substituted by a much more subtle worst-group-of-users effect, where the suggested grouping is fixed, and it is decided before the channel or the demands are known. In some cases, the theoretical gains are completely recovered, and this is done without any user selection technique. We analyze the achievable rate performance of the proposed scheme and derive insightful performance approximations which prove to be very precise.  \vspace{-0.5cm}
    \end{abstract}

	\begin{IEEEkeywords}
	    Coded-caching,  finite SNR,  shared caches, worst-user bottleneck, effective coded caching gain.\vspace{-0.5cm}
	\end{IEEEkeywords}

	\IEEEpeerreviewmaketitle
		
\section{Introduction}
	\IEEEPARstart{C}{ache}-aided communication is a promising approach toward reducing congestion in various communication networks~\cite{Ali,Paschos}. The promise of this approach was recently accentuated in the seminal paper of Maddah-Ali and Niesen~\cite{Ali}, who proposed \emph{coded caching} as a means to speed up content delivery by exploiting receiver-side cached content to remove interference. 
	
		The work in~\cite{Ali} considers the error-free (or equivalently, high-SNR) shared-link Broadcast  Channel (BC), where a transmitter with access to a library of $N$ content files serves $K$ cache-aided users. Each such user enjoys a local (cache) memory of size equal to the size of $M$ files, i.e., equal to a fraction $\gamma \triangleq \frac{M}{N} \in [0,1]$ of the library size. 
		The so-called \emph{MN scheme} of~\cite{Ali} involves a cache placement phase and a subsequent delivery phase. During the first phase, each file is typically split into a very large number of subfiles, which are selectively placed in various different caches. During the second phase, the communication process is split into a generally large number of \emph{transmission stages}, and, at each such stage, a different subset of $K\gamma+1$ users is simultaneously served via a XOR multicast transmission, thus allowing for a theoretical speed-up factor of $K\gamma+1$ as compared to the uncoded case. 
		This speed-up factor  of $K\gamma+1$ is also referred to as the Degrees of Freedom (DoF) achieved by this scheme, or similarly as the \emph{coded caching gain}.

		The above algorithm was originally developed for the scenario where the channel is error-free and the capacity to each user is identical. 
	    In recent years, a variety of works have investigated coded caching under more realistic wireless settings, considering for example uneven channel qualities\cite{Tang,Bidokhti2019_TIT,Cao2019_TC,Lampiris2020_IZS}, the role of Channel State Information (CSI) availability \cite{Jing,Zhang2015_Allerton,lampiris2017_ISIT_noCSIT,Lampiris2018_ISIT}, 
	    statistically diverse channels~\cite{Chen2017_TWC,Bayat}, and a variety of other scenarios~\cite{Lampiris2018_JSAC,lampiris2018full,Shariatpanahi,Zhong,Xu,Cao2016_TWC,Tolli2020_TWC}.
				
		Unfortunately, it is the case that coded caching suffers from two major constraints. The first is often referred to as the ``file-size constraint'' of coded caching, which, as we will recall later, effectively forces different users to fill up their caches with identical content\cite{shanmugam2016finite,Lampiris2018_JSAC}.  
		This constraint essentially foregoes the freedom to endow users with their own dedicated caches, and rather forces these users to share a very limited number of cache states that is considerably smaller than $K$. 
		On the other hand, there is a seemingly unrelated constraint which stems from the fact that the XOR multicast transmissions are fundamentally and inevitably limited by the rate of the worst user that they address\cite{Jindal2006_multicast}. 
		This constraint, often referred to as the ``worst-user bottleneck'' of coded caching, arises when users experience different channel strengths, and it is a constraint that is severely exacerbated as the ${\rm SNR}$ becomes smaller.  
		
	    Both these realities, of bounded file sizes and limited SNR, are naturally inherent to any practical wireless content-delivery system. Let us look at these bottlenecks in greater detail.

	\subsection{Subpacketization Bottleneck and the Need for Shared Caches}
		Our work builds on the premise that almost any realistic single-stream coded caching scenario will involve the use of shared, rather than dedicated, caches. As we will see right below, this has to do with the simple fact that, under realistic assumptions on $\gamma$ and $K$, the file sizes (subpacketization) required by caching schemes dwarf any realistic file sizes that we encounter in wireless downlink applications. The evidence for this is overwhelming, and, to date, under realistic assumptions, any high-performance coded caching scheme requires file sizes that grow exponentially or near-exponentially with $K$.  For example, the MN algorithm requires file sizes to be at least $\binom{K}{K\gamma}$, and as we know from~\cite[Theorem 3]{QifaPDAtit}, under some basic symmetry conditions, this same subpacketization is indeed necessary for any algorithm to achieve this same gain. 
		Similarly, it was shown in~\cite{shanmugam2016finite} that decentralized schemes  (cf.~\cite{MaddahAli2015decentralized}) require exponential (in $K$) subpacketization in order to achieve linear caching gains, and, along similar lines,~\cite[Theorem 12]{HypergraphCodedCachingTit} proved that, under basic assumptions, there exists no coded caching scheme that enjoys both linear caching gains and linear subpacketization. 
		
        Consequently, we are in a position to say that such schemes will inevitably require many users to share the same cache content. Let us consider for instance the original MN scheme. Under the constraint that file sizes cannot exceed a realistically valued $S_{max}$, we know that the best course of action is to encode over a limited number of $\Lambda<K$ users at a time, creating $\Lambda$ different cache states. This $\Lambda$ is indeed limited by the file size constraint that asks that $\binom{\Lambda}{\Lambda\gamma}\leq S_{max}$. This approach naturally limits the aforementioned (error-free) optimal gain to $\Lambda\gamma+1$ \cite{Ema}, and it entails cache replication simply because now there are only $\Lambda$ cache states to be shared\footnote{It is worth noting that this shared cache setting 
        not only captures the effect of the file-size constraint, but also reflects promising heterogeneous scenarios where a main station serves users with the help of smaller cache-endowed helper stations~\cite{Golrezaei2012,Ema}.} among the $K$ users. 
        As we will show later on, this forced replication can be exploited to circumvent another major problem: the worst-user bottleneck.

	\subsection{Worst-User Bottleneck: Motivation, Nature of the Problem, and Prior Work}
		As we have mentioned, the worst-user limitation induced by the nature of the multicast transmission\cite{Jindal2006_multicast} is exacerbated when ${\rm SNR}$ becomes smaller and when the channel strengths are different.  
		Consequently, this dependence on multicasting can severely affect the applicability of coded caching in many wireless scenarios that possess such characteristics.  
		Such scenarios prevail in cellular or satellite communications settings~\cite{Jia} that suffer from heavy path-loss and/or shadowing,
		and in IoT networks 
		\cite{Nolan2016} or massive Machine‐Type Communication (mMTC) settings~\cite{Bockelmann2016}.  
		Similarly, we know that in $4$G LTE networks the range of users' signal-to-interference-plus-noise ratio (SINR) is typically $0$--$20$ dB \cite{Teltonika}, 
		while the SINR of cell-edge users can be closer to $0$--$5$~dB.  			
		The worst-user bottleneck is also exacerbated when considering the well-established setting of quasi-static fading that we will consider in the following, and which generally comes about in the presence of longer coherence periods and shorter latency constraints. 
		This quasi-static setting applies to low-mobility scenarios which nicely capture coded-caching use-cases where pedestrians or static users consume video streaming. 
	
		This bottleneck has sparked considerable research interest that resulted in a variety of notable results~\cite{Tang,Yang2016_ISTC,Ngo,Tegin2020,Daniel2020}. 
		For example, the work in~\cite{Ngo} shows that,  
		in a single transmit-antenna setting with finite power and quasi-static fading, the effective gain does not scale as~$K$ becomes larger even in the absence of a file-size constraint; moreover, the power must scale linearly with~$K$ in order to preclude the collapse of the multicast rate (cf.~\cite[Table~I]{Ngo}). 
		Taking a different approach, the work in~\cite{Daniel2020} employs superposition coding for opportunistic scheduling. Another notable work can be found in~\cite{Tegin2020}, which groups together users that experience similar channel (similar SNR), and which, after neglecting  users with the weakest channels, delivers to each group separately. To date, for the single transmit-antenna setting without user selection, no scheme is known to overcome the worst-user bottleneck. 
    	
\begin{figure}[t]
			\setlength{\abovecaptionskip}{4pt}
			\setlength{\belowcaptionskip}{0pt}
			\centering
			\includegraphics[width=3in]{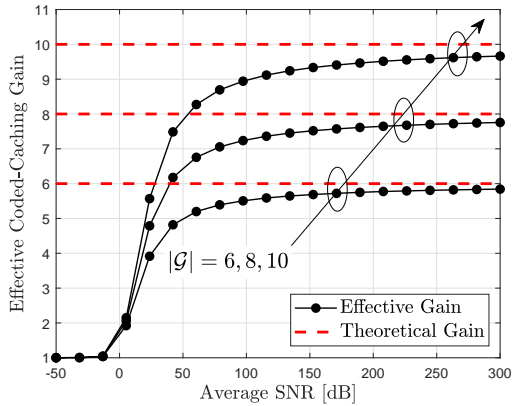}
			\caption{Ratio between the average rates of the MN scheme and TDM (i.e., the \emph{effective coded-cahing gain}) over quasi-static Rayleigh fading channel for different values of $|\Gc|=\Lambda\gamma+1$.}\vspace{-0.5cm}\label{Gain_MN_intro_fig}
		\end{figure}

    In all these scenarios, this bottleneck essentially diminishes the aforementioned coded caching gain\footnote{We remind the reader that the gain describes the cache-aided speed-up factor over the uncoded approach which employs the basic Time Division Multiplexing (TDM) method that serves one user at a time.}. Had the SNR been infinite, or the instantaneous link strengths identical, this hypothetical gain would have taken the form $|\Gc|=\Lambda \gamma+1$, for any allowable $\Lambda$ up\footnote{We also remind the reader that this allowable $\Lambda$ is generally much less than $K$, due to the bounded file sizes.} to $K$.  
		Yet, as the SNR decreases, the effect of the worst-user bottleneck becomes more accentuated\footnote{To see this, simply recall that for smaller values of SNR and for $z<1$, then $\ln(1+z\text{SNR})\approx z\ln(1+\text{SNR}) $.}, and the effective gain eventually collapses. 
		This collapse will be rigorously described in Prop.~\ref{prop:collapse_caching_gains}, and it is illustrated in Fig.~\ref{Gain_MN_intro_fig}.

	\subsection{Contributions and Organization} 
		In this work, we consider coded caching with centralized placement in the standard single-antenna BC, in the context of finite SNR and quasi-static fading. 
		The analysis holds for any SNR, and some of the subsequent approximations either imply many users, or imply lower SNR values. Both these asymptotic regimes manage to crisply and very precisely characterize the performance, and both govern modern wireless communications. Our contributions are outlined as follows.

			We first show that the coded caching gain of the MN scheme with respect to simple uncoded TDM (either with or without file size constraints) deteriorates considerably for any reasonable range of SNR values, and, in fact, completely vanishes in the low-SNR regime. 
			
			Then, focusing on the file-size constrained scenario (which corresponds to having a limited number $\Lambda$ of different cache states), we present a novel transmission scheme that substantially improves the gain, and which manages to recover -- without any user selection -- the entire theoretical coded-caching gain $\Lambda\gamma+1$ in the presence of sufficiently many~users. 	The proposed scheme, which will be referred to as the \emph{Aggregated Coded-Caching  scheme}, builds on the inevitability of having users with identical cache content, and it employs multi-rate encoding that avoids XOR transmissions, thus allowing each user to receive at a rate that matches its single-link capacity. 
			In fact, it turns out that having $B$ users per cache state is as efficient as having a time diversity of $B$ coherence times. 
			
		    We analyze the average rate (which we rigorously define later) and derive its exact analytical expression. 
		    To offer insight, we apply low-SNR approximations, as well as large-$K$ approximations, to derive clear closed-form expressions for the average rate and the gain. These approximations are shown to retain a robust accuracy even for a very modest user count. 
		As a consequence of these results, we now know a simple way to exploit the unavoidable nature of the file-size constraint in order to almost entirely remove the worst-user bottleneck. In essence, we show that, given the file-size constraint, the worst-user effect can be made negligible. 

		The remainder of this paper is organized as follows: Section~\ref{se:sysmod} defines the system model and the problem considered. In Section~\ref{se:bs_scheme}, the proposed scheme is formally described. 
		The average rate is investigated in Section~\ref{se:ear_analysis}, where we derive several tight approximations. Some numerical results and comparisons are presented in Section~\ref{se:num_analysis}, and Section~\ref{se:conclusions} concludes the paper.  

	\subsubsection*{Notation}
	    We use the notation $X\sim \Yc$ to state that a random variable $X$ follows the distribution~$\Yc$. 
	    Given a real-valued function $f(x)$ over a variable $x$, $f(x) = o(x)$ stands for $\lim_{x\to 0}\frac{f(x)}{x} = 0$.
		$\E\{\cdot\}$ denotes the expectation operator.
		We use the short-hand notation $[n]\triangleq \{1,2,\dotsc,n\}$ for a positive integer $n$. 
		$\mathcal{N}(\mu,\sigma^2)$ denotes the Gaussian distribution of mean $\mu$ and variance $\sigma^2$. 
		$|\cdot|$ denotes the cardinality operator of a set. 
		All sets are assumed to be ordered. 
		
\section{System Model and Problem Definition}\label{se:sysmod}

 	 We consider the quasi-static Rayleigh fading BC in which a single-antenna transmitter serves a set of $K$ users. As mentioned before, each user requests a file from a library ${\mathcal{F}}=\{W_n\}_{n=1}^N$ of $N$ files, and each user is assisted by a cache of normalized size $\gamma \in [0,1]$. We will consider an arbitrary number $\Lambda$ of different allowable cache states, and we will assume for simplicity that $K$ is an integer multiple of $\Lambda$.

	The received signal at user~$k \in [K]$ is given by $Y_k = H_k X + Z_k$, where $H_k$ denotes the channel coefficient for user~$k$, $X$ denotes the transmit signal satisfying an average power constraint $\E[|X|^2]\leq P$, and $Z_k$ denotes the zero-mean, unit-power, additive white Gaussian noise at user~$k$. Each user~$k$ experiences an instantaneous SNR of $\SNR_k = P |H_k|^2$, and an average ${\rm SNR}$ of $\rho \triangleq \E_{H} \left\{ {\rm SNR}_k \right\}$. As is common in the coded caching literature (cf.~\cite{Ngo}), we will assume that $H_k$ remains fixed during a transmission stage, but may change between different transmission stages. We will further assume that the users experience statistically symmetric Rayleigh fading. 
	
	As with various other works that study coded caching under quasi-static fading~\cite{Tang,Ngo,Yang2016_ISTC}, we will adopt the transmission rate\footnote{We recall that, for quasi-static Rayleigh fading, the typical metric of the worst-case \emph{delivery time} does not have an expectation.} as the metric of interest. Toward this, we define the instantaneous rate $r_k$ as the maximal rate that can be transmitted to user~$k$ for the instantaneous channel realization. Similarly, we will consider the \emph{average rate} $\E_{H}\{r_k\}$ to be the average -- over the fading statistics -- of the above instantaneous rate. 
	It is important to not confuse this long-term average $\E_{H}\{r_k\}$ with the ergodic rate, which implies an ability to encode over several fading realizations (cf.~\cite{Yang2016_ISTC,Ngo}). 

    In this context, a coded caching scheme seeks to provide an \emph{effective coded-caching gain},
    which represents the true (multiplicative) speed up factor, at finite SNR, that the said scheme offers over the average rate obtained by TDM.  
    This effective gain is contrasted to the (ideal, or high-SNR) \emph{nominal coded-caching gain}, which is the gain $\Lambda\gamma+1$ provided by the file-size constrained coded caching in the error-free scenario with fixed and identical link~capacities. 
    
    The proposed scheme and the analysis are motivated by the fact that the effective gain of the MN scheme collapses at low SNR,  which will be proven in Section~\ref{se:ear_analysis}. 
    This collapse will be irrespective of $\Lambda$ and $K$, i.e., it happens even in the absence of file-size constraints.

	\section{Aggregated Coded-Caching Scheme}\label{se:bs_scheme}
	In the following, we introduce a novel scheme, coined as the \emph{Aggregated Coded Caching} (ACC) scheme, which will be shown to overcome the previous collapse of the effective gains. 

    The scheme clusters the users into $\Lambda$~\emph{groups} of $B = K / \Lambda$ users per group, such that every member of the same group is assigned identical cache content (i.e., employs shared caches). As we have seen, this is essentially inevitable under realistic file-size constraints. The scheme also follows a standard clique-based approach~\cite{Ali}, such that the transmission is divided into \emph{transmission stages} that experience a clique-side information pattern. This means that, as in~\cite{Ali}, for each such stage, any desired subfile of some served user can be found in the cache of every other user involved in that same transmission stage. Thus, this approach defines a side-information structure that was addressed in the following well known result from~\cite{Tuncel2006}.

		\begin{prop}[{\!\cite[Thm. 6]{Tuncel2006}}]\label{prop:cap_reg_bc}  
			The capacity region of 
			a $t$-user Gaussian~BC, where each user $i\in[t]$ is endowed with SNR equal to~$\SNR_i$ and requests message $W'_i$ while having access to side information $\overline\Wc_i =\! \{W'_j\}_{j\neq i, j \in [t]}$, is given by
				\begin{align}
					\Cc = \big\{(R_1,\cdots,R_t) : 0 \leq R_i \leq \log_2(1+ \SNR_i),\ i\in[t]\big\}.\notag
				\end{align}
		\end{prop}		
		
		\begin{proof}
			Proposition~\ref{prop:cap_reg_bc} is known as a special case of~\cite[Thm. 6]{Tuncel2006} and this particular form has been considered in~\cite{Kramer2007_ITW,Asadi2015_TIT}. 
			More details on this, as well as on the association to our setting, are described in Appendix~\ref{app:capacity_region}.   
		\end{proof}

		Proposition~\ref{prop:cap_reg_bc} implies that, under this particular configuration of side information, each user can achieve its own point-to-point capacity, as if no other user was being served at the same time. 
		There are various optimal \emph{multi-rate transmission} schemes for this setting~\cite{Kramer2007_ITW,Asadi2015_TIT}, and the proposed ACC scheme can remain oblivious to the encoding choice.\footnote{In terms of practicality, it is known that very simplified schemes, such as nesting BPSK into M-QAM constellations (cf.~\cite{Tang}), come extremely close to achieving the above capacity region, and in fact achieve the single-user capacity insofar as we restrict ourselves to QAM modulations~\cite{Xue2007}.    
			If necessary, such simplified codes can be directly applied in our cache-aided setting, with only minor performance losses.} 		 
	
	\begin{remark} \label{remark_2_clique}
    	We state in advance that the aforementioned multi-rate transmission must indeed be combined with the method of shared caches in order to yield the desired gains. 
    	While multi-rate transmission performs better than MN-based XORs, this rate improvement appears only when we focus our attention on a single isolated delivery stage that serves some fixed set of users $\Gc$. However, when considering the entire delivery problem over all sets $\Gc$, we would see no gain because the MN placement and multicast group generation without shared caches would not allow for an additional subfile to be sent to a potentially `fast' user in~$\Gc$, without generating interference to the remaining (slower) users. This latter point, which is that the MN placement does not allow exploitation of fast users, is presented below in the original context of XORs. 
	\end{remark}
    \begin{example}\label{example_2_clique}
        Consider delivery of XOR $ A_{2,3}\oplus B_{1,3}\oplus C_{1,2}$ meant for users $\Gc = \{1,2,3\}$ who respectively ask for files $W_1 = A, W_2 = B, W_3 = C$. Even if user 1 decodes $A_{2,3}$ very quickly, she must wait for $B_{1,3}$ and $C_{1,2}$ to be decoded, because -- by definition of the MN placement -- there exists only one subfile that is desired by user 1 and which can be decoded by users 2 and 3. 
    	An illustrative example is represented in Fig.~\ref{fig:scheme_no_sharedcaches}. 
    \end{example}

	\subsection{Aggregated Coded-Caching Design}
		We proceed with the description of the placement and delivery phases of the ACC scheme. At the end, we will also present a small clarifying example.

		\subsubsection{Placement Phase}
			This phase begins by arbitrarily splitting the $K$ users into $\Lambda$ ordered groups of $B = \frac{K}{\Lambda}$ users~each.  Placement is exactly as in~\cite{Ema}, and thus it simply applies the MN placement of the $\Lambda$-user problem, such that each user of the same group shares the same cache content. In particular, each file $W_n$, $n\in[N]$, is partitioned into ${\Lambda \choose \Lambda\gamma}$ segments as
				$W_n\!\to\! \left\{ W_n^\mathcal{T} : \mathcal{T}   \subseteq[\Lambda],\, |\mathcal{T}|= \Lambda\gamma\right\},$
			and then each user in group~$g\in[\Lambda]$ stores all the subfiles in the following set
				$\Zc_g = \{W_n^\Tc : \Tc \subseteq [\Lambda],\, |\Tc| = \Lambda\gamma,\, \Tc \ni g,\, \forall n\in[N] \}.$

				\begin{figure}[t]
					\begin{flushleft}
						\begin{subfigure}[t]{.465\textwidth}
							~\hspace{-2ex}\includegraphics[width=3.2in]{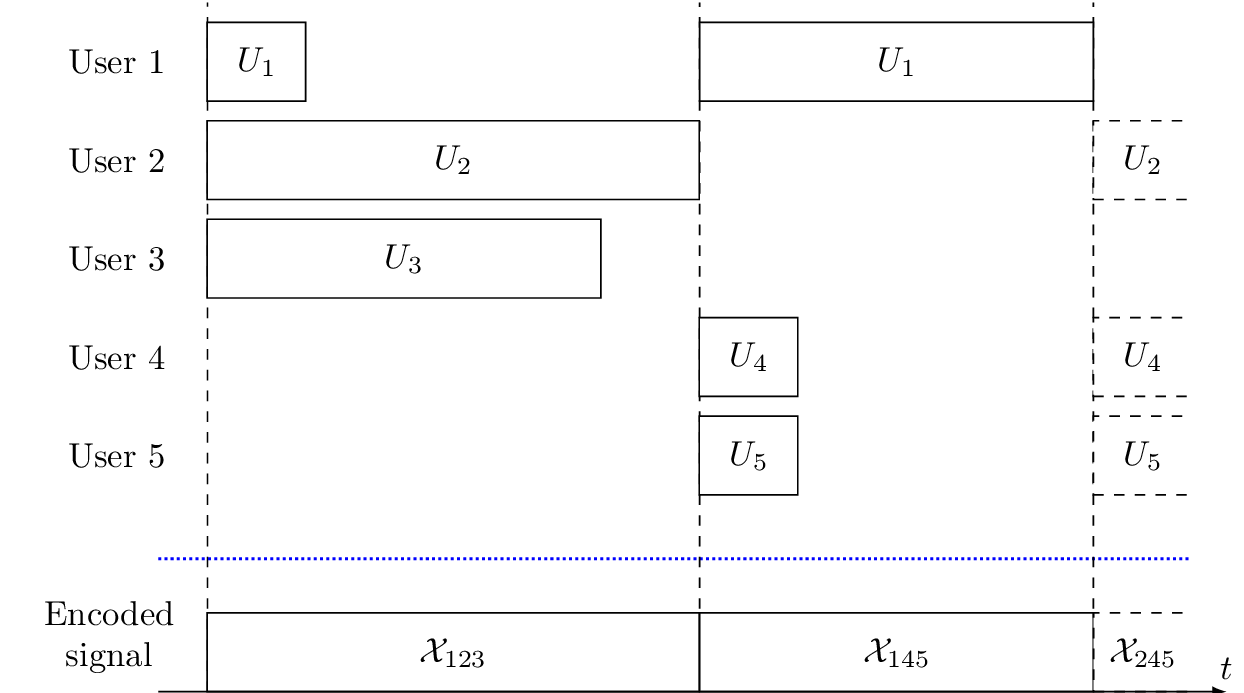}%
							\caption{\small Dedicated caches: The total delay depends on the worst-user capacity at each transmission stage.  $\Xc_{abc}$ denotes the signal encoded for the users $a$, $b$, and $c$.}
							\label{fig:scheme_no_sharedcaches}
						\end{subfigure}\hspace{0.03\textwidth}
						\begin{subfigure}[t]{.495\textwidth}
								~\hspace{-2ex}\includegraphics[width=3.4in]{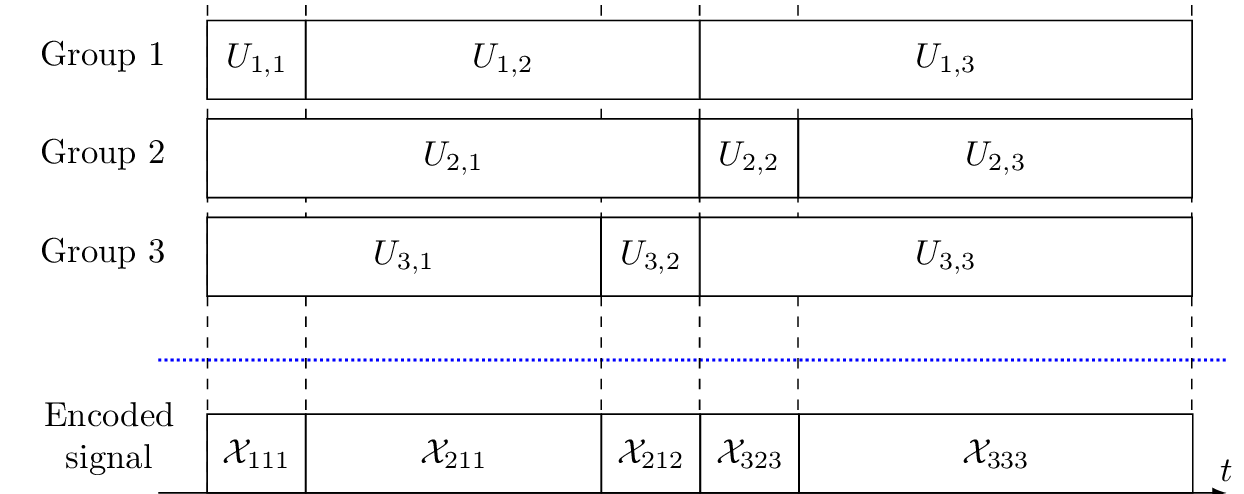}%
								\caption{ACC scheme: Delay depends on the average per-user rate within the group. 
								$\Xc_{abc}$ denotes the encoded signal  for users $a$, $b$, and $c$ of groups $1$, $2$, and $3$,~respectively.}	
								\label{fig:scheme_sharedcaches}
						\end{subfigure}
					\end{flushleft}~\vspace{-3ex}
					\caption{Comparison of MN and ACC schemes for a nominal coded-caching gain of $3$. 
					}\vspace{-0.5cm}\label{fig:comparison_encoding}
				\end{figure}

		\subsubsection{Delivery Phase}

			The delivery phase is split into ${ \Lambda\choose \Lambda\gamma+1}$ transmission stages, where each stage involves a set $\Gc\subseteq [\Lambda]$ of $|\Gc|=\Lambda\gamma+1$ groups. 
			During each stage, the transmitter \emph{simultaneously} delivers to as many as  $|\Gc|=\Lambda\gamma+1$ users, each from a different group in set~$\Gc$. The users within each group are served one after the other in a round-robin manner. 
			For a given set~$\Gc$, the transmitter employs a multi-rate code that achieves the capacity in Proposition~\ref{prop:cap_reg_bc}. 
					
			We will use the notation $\Xc\big(A_{1},\dotsc,A_{|\Gc|}\big)$ to represent the transmit signal when the transmitter delivers the subfiles $A_{1},\dotsc,A_{|\Gc|}$. 
			When describing the delivery, we will use vector~$\bv\in{\Zb}^{|\Gc|}$ to represent the set of users that are being served at a particular time\footnote{Please note here that the dependence of $\bv$ on the time index and on $\Gc$ is assumed but omitted for simplicity.}. 
			Furthermore, let $\Gc(i)$ denote the $i$-th group in $\Gc$, $i\in[|\Gc|]$ (recall the group-set $\Gc$ is ordered); consistently, $\bv(i)$ tells us which user of the $i$-th group in~$\Gc$ is currently being served, where $\bv(i)\in[B]$. 
			The scheme will serve the users $\bv$ of the groups in $\Gc$ by transmitting 
				\begin{align}\label{eq:encoded_signal_bs}
					X_{\Gc,\bv}^{} = \Xc\big(\big\{W_{d_{\bv(i)}}^{\Gc \setminus \{\Gc(i)\}} \big\}_{i\in[|\Gc|]}\big),			
				\end{align}
			where $d_{{\bf v}(i)} \in [N]$ denotes the file index requested by user ${\bf v}(i)$.
			Algorithm~\ref{alg:bs_transmission} presents the transmission for a specific set $\Gc$ of groups. Every time the user of some group $\Gc(i^\prime)$ obtains its subfile, $\bv(i^\prime)$ is updated\footnote{We are actually incurring an abuse of notation in~\eqref{eq:encoded_signal_bs} and Algorithm~\ref{alg:bs_transmission}. 
			Specifically, when a group updates its served user, the transmitter continues encoding the partially-decoded subfiles taking into account that there  remains only a part of such subfiles to be transmitted. 
			This is intuitive from Fig.~\ref{fig:scheme_sharedcaches}. 
			}
			 as $\bv(i^\prime) \leftarrow \bv(i^\prime) + 1$. 			
		
			This process is repeated until all the users in all the groups in $\Gc$ are served. 
			If every user of a group has obtained its intended subfile, the transmission can be composed only of the remaining groups. 
			Algorithm~\ref{alg:bs_transmission} is iterated over all possible ${\Lambda \choose \Lambda\gamma + 1}$ sets $\Gc$. After this,  the $K$ users obtain their requested files. We reemphasize that the ACC scheme does not apply user selection. 
			Let us proceed with a simple clarifying example.
				\begin{algorithm}[t]\caption{Transmission stage for a set of groups~$\Gc$}\label{alg:bs_transmission}
					\DontPrintSemicolon
					\SetKwRepeat{Transmit}{Transmit}{until}\SetKw{Init}{Initialize}
					\newcommand\mycommfont[1]{\footnotesize \ttfamily\textcolor{gray}{#1}}\SetCommentSty{mycommfont}
						\Init $\bv\in\Zb^{|\Gc|}$ as $\bv(i) \longleftarrow 1$ for any $i\in[|\Gc|]$\\		
						\Init $\mathrm{Number~of~finished~groups} \longleftarrow 0$ 
						
						\While{$\mathrm{Number~of~finished~groups} \neq |\Gc|$}{
							\Transmit{A served user $\bv(i)$, $i\in[|\Gc|]$, fully obtains its subfile 
							}{$X_{\Gc,\bv}  \longleftarrow  \Xc\Big(\Big\{ W_{d_{\bv(i)}}^{\Gc \setminus \{\Gc(i)\}} \Big| i\in[|\Gc|] \text{ and } \bv(i) \le B \Big\}\Big)$
							}
							Set $i^\star$ as the index of the group $\Gc(i^\star)$ whose user has decoded its subfile \\
							\If{$\bv(i^\star) = B$}{$\mathrm{Number~of~finished~groups}  \longleftarrow \mathrm{Number~of~finished~groups}  + 1$}
							$\bv(i^\star) \longleftarrow \bv(i^\star) + 1$
						}
				\end{algorithm}

				\begin{example}\label{ex:example_simple}
					Consider a transmission stage that serves groups $\{1,2,3\} = \Gc$, where each group is composed of $B = 3$  users. To simplify the explanation of this example, let us denote the $b$-th user of the (ordered) group $g$ as~$U_{g,b}$, and let $W'_{g,b}$ denote the subfile intended for this user. 
					Let us further assume that the normalized capacity of each user (expressed in transmitted subfiles per time slot) is as in the next table\vspace{-0.15cm}
						\begingroup\renewcommand*{\arraystretch}{1.2}
							\begin{table}[h]\centering
								\begin{tabular}{ c|c|c|c } 
									& User 1 & User 2 & User 3\\
									\hline
									Group 1 & 1 	 & 0.25 & 0.2 \\ 
									Group 2 & 0.2	 & 1 		& 0.25 \\ 
									Group 3 & 0.25 & 1 		& 0.2    
								\end{tabular}
							\end{table}\vspace{-0.5cm}
						\endgroup	\FloatBarrier
								
					\noindent which simply implies that the point-to-point capacity of users $U_{1,1}$, $U_{2,2}$, and $U_{3,2}$ is four times the capacity of users $U_{1,2}$, $U_{2,3}$, and $U_{3,1}$, and five times the capacity of $U_{1,3}$, $U_{2,1}$, and $U_{3,3}$. 
					The encoded signal for this example is illustrated in Fig.~\ref{fig:scheme_sharedcaches}. 
					Initially, the first user of each group is selected to be served, and the transmitter sends $\Xc\big(W'_{1,1},W'_{2,1},W'_{3,1}\big)$. Following the result of Proposition~\ref{prop:cap_reg_bc}, each user can decode its own subfile at a rate matching its single-user capacity ($\log_2(1+\SNR_{g,b})$) because each user knows the subfiles of the other two served users. 
									
					After the first slot, user $U_{1,1}$ has successfully decoded its subfile. Hence, $U_{1,1}$ is substituted by $U_{1,2}$, and the transmitter sends $\Xc\big(W'_{1,2},W'_{2,1},W'_{3,1}\big)$. The key is that we can serve any of the users sharing the cache because all of them can cache out the subfiles intended by the users of the other groups in~$\Gc$, and vice versa. 
					Thus, every time a user  obtains its subfile, a new member of the same group substitutes this user. After the fourth time slot, $U_{3,1}$ obtains its subfile, and is replaced by $U_{3,2}$, so the transmitter then sends $\Xc\big(W'_{1,2},W'_{2,1},W'_{3,2}\big)$. After the fifth slot, each of the three served users obtains its desired subfile and the transmitter begins to send $\Xc\big(W'_{1,3},W'_{2,2},W'_{3,3}\big)$, and so on.					
				\end{example}

    \section{Average Rate Analysis}\label{se:ear_analysis}
    
    	In this section, we analyze the long-term average rate of the MN and ACC schemes. 
    	First, we will derive the exact expression of the average rate for both schemes. 
    	Afterward, we will approximate this rate at low SNR, and we will also derive the limit in the regime of many users. 
    	It will turn out, as we will see in the following, that these two approximations are very robust in realistic scenarios. Furthermore, we obtain the effective gain of this scheme with respect to TDM as well as its improvement with respect to the MN scheme, and we show that while the effective gain of the MN scheme vanishes at low SNR, the ACC scheme recovers -- at any SNR value -- the nominal (high-SNR) gain as the number of users per cache increases. 
    	
    	We recall that, under Rayleigh fading,  the SNR follows an exponential distribution. 
    	Hence, for user~$k \in [K]$, the probability density function (PDF) and cumulative distribution function (CDF) of $\SNR_k$ are given respectively by
    	      $f_{{\rm SNR}_{k}}(x) = \frac{1}{\rho} \exp\left(-\frac{x}{\rho}\right)$ 
    	   and
    	      $F_{{\rm SNR}_{k}}(x) = 1-\exp\left(-\frac{x}{\rho}\right)$, for any $x \ge0$, 
	    where $\rho = \mathbb{E}_H\{{\rm SNR}_k\}$ denotes the average SNR with respect to channel states (recall that the users’ channels are statistically symmetric).
    	As for the ACC scheme, we will use $\SNR_{g,b}$, $f_{{\rm SNR}_{g,b}}(x)$, and $F_{{\rm SNR}_{g,b}}(x)$ to refer to the SNR, PDF, and CDF corresponding to the $b$-th user of the group~$g$, where $b\in[B]$ and $g\in[\Lambda]$.

        \subsection{Average Rate of the MN and ACC Schemes}\label{se:aver_scheme}
    
            \subsubsection{Average Rate of the MN Scheme}\label{se:mn_scheme}
        
                We first note that, since the MN scheme is designed for the setting with dedicated caches, in a setting with $\Lambda$ cache states and $B = K/\Lambda$ users per cache, the transmission consists of repeating $B$ times the transmission of the dedicated caches setting. 
                Consequently, the MN scheme consists of  $B\binom{\Lambda}{\Lambda\gamma+1}$ transmission stages, each of them employed to deliver a XOR to a group of users of size $|\Gc| = \Lambda \gamma+1$.
                
        		Consider the delivery to a particular set $\Gc$ of $\Lambda\gamma+1$ users. 
				We know from the multicast capacity theorem in~\cite{Jindal2006_multicast} that the maximum instantaneous rate for any user~$i\in\Gc$ takes the form
            		\begin{align}\label{eq:inst_rate_def2}
            			r_{i,\Gc}^\emn = \log_2 \Big(1+\min_{k\in\Gc}\SNR_k \Big) \quad {\rm bits/s/Hz}. 
            		\end{align} 
                Hence, the instantaneous sum rate is given by $\sum_{i\in\Gc}r_{i,\Gc}^\emn$, since we are simultaneously serving all the $|\Gc|$ users. 
				Consequently, the average sum rate for that specific set $\Gc$ takes the form
            	    \begin{align}
        			   \rmn_{\Gc}  \triangleq  \E_{H} \Big\{ \sum\nolimits_{i\in\Gc}r_{i,\Gc}^\emn \Big\} 
        				     = \frac{|\Gc| }{\ln 2}\E_{H} \big\{ \ln(1+\min\limits_{k \in \mathcal{G}} {\rm SNR}_{k}) \big\} \label{EAR_MN_Intro1end}
        			\end{align}	
                which follows because the users are statistically equivalent, which in turn also implies that the average sum rate $\rmn$ remains the same for any set $\Gc$, i.e., it implies that $\rmn = \rmn_{\Gc'}$ $\forall\Gc'\subseteq [\Lambda],\ |\mathcal{G'}| = \Lambda \gamma+1$.

				Naturally, the average rate under the TDM scheme, which we denote as $\bar R^{\rm (TDM)}$, is a special case of $\bar R^{\rm {(MN)}}$ obtained by setting $|\Gc|=1$. 
				By taking into account that $\min_{g \in \mathcal{G}}\{{\rm SNR}_{g}\}$ follows an exponential distribution with rate $\nicefrac{|\Gc|}{\rho}$, it follows from~\cite[Eq. (15.26)]{Slim_book} that 
					\begin{align}
						\Bar{R}^{\rm (MN)} &=  -\frac{|\Gc|}{\ln 2} \exp\left(\frac{|\Gc|}{\rho}\right) \cdot {\rm Ei}\left(-\frac{|\Gc|}{\rho}\right),\label{EAR_MN_Intro} 
					\end{align}
						where ${\rm Ei}(\cdot)$ represents the exponential integral function~\cite{Gradshteyn}. Note that $|\Gc|=1$ in \eqref{EAR_MN_Intro} yields the closed-form expression for $\Bar{R}^{\rm (TDM)}$.

				\subsubsection{Average Rate of the ACC Scheme}\label{se:acc_scheme}
					Due to the symmetry of the ACC scheme and the statistical symmetry of the channel, we will here focus on a particular set $\Gc$ of $|\Gc|=\Lambda\gamma+1$ user groups, where we recall that each group is composed of $B$ users. 
            	
        	As explained in Section~\ref{se:bs_scheme}, the ACC scheme allows us to serve some user~$b$ of group~$g$ at its own point-to-point capacity, and it allows us to immediately start serving another user of the same group as soon as the said  user~$b$  has completed the decoding of its subfile. 
          Furthermore, in the ACC scheme, the delivery to a group-set $\Gc$  is completed when every user belonging to one of these groups has obtained its subfile.  Consequently, the (per-user average) rate with which any group~$j$ in the set~$\Gc$ is served is here captured by             
         		\begin{align}\label{eq:inst_rate_acc1}
        			r_{j,\Gc}^\eacc = \min_{g\in\Gc}\frac{1}{B}\sum\nolimits_{b=1}^B\log_2(1+\SNR_{g,b}) \ \ {\rm bits/s/Hz}, \quad \forall j\in\Gc. 
            \end{align}         	  
          This metric is not exactly the inverse of the delay, but, since the delay has no expectation for this setting, this average rate is very useful because it crisply reflects the worst-user effect. 
        	By applying the same reasoning as in~\eqref{eq:inst_rate_def2}--\eqref{EAR_MN_Intro1end}, we obtain that the average rate with which the transmitter delivers data across the users is given by
        		\begin{align}\label{EAR_ACC_Intro1d}
        			\Bar{R}^\eacc \!=\! \frac{|\Gc|}{\ln 2} \E_H \Big\{ \min_{g\in\Gc}\frac{1}{B}\sum\nolimits_{b=1}^B\ln(1+\SNR_{g,b}) \Big\} \ \   {\rm bits/s/Hz}.
        		\end{align}	
          We quickly note that, by comparing~\eqref{EAR_ACC_Intro1d} with~\eqref{EAR_MN_Intro1end}, we can see how the worst-user effect is essentially averaged out into a cumulative ``worst-group'' effect. 
          By considering dedicated caches (i.e., by setting $B=1$), we obtain the same average rate as that of the MN scheme in~\eqref{EAR_MN_Intro1end} despite having a different (not XOR-based) coding scheme. 
        
        	In the following, $\jmath\triangleq\sqrt{-1}$ denotes the imaginary unit, ${\rm Im}\{\cdot\}$ the imaginary part of a complex number, and ${\rm E}_{-\jmath t}(\cdot)$ denotes the exponential integral function of the $(-\jmath t)$-th order~\cite{Gradshteyn}. 
        		
        		\begin{lemma}\label{lem:exact} 
        			The exact average rate of the ACC scheme over symmetric quasi-static Rayleigh fading can be derived in a double-integral form as follows
        				\begin{align}\label{AER_BS_Int}
        					&\Bar{R}^{\rm (ACC)}=\frac{|\Gc|}{B \ln2} 
        					\int\limits_0^\infty \left( \frac{1}{2} +\frac{1}{\pi} \int\limits_0^\infty \frac{{\rm Im} \left\{ \exp({-\jmath x t}) \frac{\exp({B / \rho})}{\rho^B} {\rm E}_{-\jmath t}^B \left(\frac{1}{\rho}\right) \right\}}{t} \dd t \right)^{|\Gc|} \dd y.
        				\end{align}
        		\end{lemma}
        
        		\begin{proof}
        			The proof is relegated to Appendix \ref{app:proof_lemma_exact}.
        		\end{proof} 
        		The numerical implementation of the above expression is very complex and it provides little insight. 
        		In the following, we obtain the effective gains in both the low-SNR limit and the large-$B$ limit for the MN scheme and the ACC scheme, and we derive approximations of their rates,
        		from which some meaningful insights can be easily drawn.

    \subsection{Rate Approximations and Effective Gains at  Low SNR}\label{se:low_snr}

		\subsubsection{MN Scheme}
			First, we present a low-SNR approximation for the average rate of the MN scheme, which is in fact  a special case of the ACC scheme with  $B=1$.
			Although the exact form has been derived in \eqref{EAR_MN_Intro}, we can provide a simple but tight approximation which allows us to remove the special function ${\rm Ei}(\cdot)$ from the expression. 
			\begin{lemma}\label{lem:ear_mn_approxLow}   
				In the low-SNR region, the average rate of the MN scheme can be  approximated by
					\begin{align}\label{EAR_Robustapprox}
						\Bar{R}^{\rm (MN)} \approx \frac{|\Gc|}{\ln 2} \left(   \ln\left(1+\frac{\rho}{|\Gc|}\right)-\frac{\rho^2}{2 |\Gc|^2 \left(1+\rho / |\Gc|\right)^2}  \right).
					\end{align}
			\end{lemma}

			\begin{proof}
					See Appendix~\ref{app:ear_mn_approxLow}.
			\end{proof}

			In the numerical evaluation section (see Fig.~\ref{EAR_fig_com} in Section~\ref{se:num_analysis}), it will be shown  that this computationally efficient second-order approximation  
			can in fact provide us with an extremely reliable estimation of the performance even in the medium-SNR region.
			
			Let us now consider the exact effective gain of the MN scheme, which -- directly from~\eqref{EAR_MN_Intro} -- takes the form
				\begin{align}\label{Omega_MN_exact}
					\frac{\Bar{R}^{\rm (MN)}}{\Bar{R}^{\rm (TDM)}} & =  \frac{|\Gc| \exp\left(\frac{|\Gc|}{\rho}\right) \cdot {\rm Ei}\left(-\frac{|\Gc|}{\rho}\right)}{ \exp\left(\frac{1}{\rho}\right) \cdot {\rm Ei}\left(-\frac{1}{\rho}\right)}.
				\end{align}
			As expected, the effective gain converges to the nominal gain $|\Gc|$ at high SNR,  since the limit of~\eqref{Omega_MN_exact} as $\rho\to\infty$ is  $|\Gc|$. 
			On the other hand, in the low-SNR region, this effective gain entirely vanishes, as stated in the following proposition. 
			
			\begin{prop}\label{prop:collapse_caching_gains}
				For any value of $K$ and $\Lambda$, the effective gain of the MN scheme converges to  
						\begin{align}
						\lim_{\rho\to 0} \frac{\Bar{R}^{\rm (MN)}}{\Bar{R}^{\rm (TDM)}} = 1
						\end{align}
							meaning that this effective coded-caching gain entirely vanishes at low~SNR.
			\end{prop}	 	  
			\begin{proof}
					See Appendix~\ref{app:gain_mn_approxLow}.
			\end{proof}	

			As noted before, Proposition~\ref{prop:collapse_caching_gains} holds for any scheme which requires decoding of single XORs.

		\subsubsection{ACC Scheme}
			Let us now consider the ACC scheme. In the following, for any integer vector $\bb\triangleq[b_1,b_2,\cdots,b_B]\in\Zb^{B}$ composed of $B$ non-negative elements, we will use ${n \choose {\bb}}\triangleq \frac{n!}{b_1!b_2!\cdots b_B!}$ to denote the multinomial coefficient. We can now state our following result.  
					
			\begin{lemma}\label{lem:ear_approxLow} 
				In the low-SNR region, the average rate of the ACC scheme can be approximated by
					\begin{align}\label{AER_Linear_approx}
						\racc &\approx  \frac{\rho |\Gc|}{B \ln2}\, \Psi_{|\Gc|},
					\end{align}
				since it holds that $\racc =   \frac{\rho |\Gc|}{B \ln2}\, \Psi_{|\Gc|}+o(\rho)$, where $\Psi_{|\Gc|}$ is defined as
					\begin{align}
						\Psi_{|\Gc|}  \triangleq  \sum\nolimits_{||{\bf b}||_1=|\Gc|} {|\Gc| \choose {\bf b}} \frac{|\Gc|^{-1-\sum_{t=1}^B (t-1)b_t}}{\prod_{t=1}^{B} ((t-1)!)^{b_t}} \Big(\sum\nolimits_{t=1}^B (t-1)b_t\Big)!\ , 
					\end{align}
				and where the summation is over all the vectors composed of $B$ non-negative integer elements and whose norm-1 equals $|\Gc|$. 
			\end{lemma}

			\begin{proof}
				The proof is relegated to Appendix~\ref{app:proof_approx_lowSNR1}.
			\end{proof} 
			
			From  Lemma~\ref{lem:ear_approxLow}, we obtain the following corollary.
			\begin{coro}\label{Delta_coro}
				In the limit of low SNR, the ratio of $\Bar{R}^{\rm (ACC)}$ over $\Bar{R}^{\rm (MN)}$ converges to the constant 
					\begin{align}\label{eq:approx_lowSNR_cor}
						\lim_{\rho\to 0}\frac{\Bar{R}^{\rm (ACC)}}{\Bar{R}^{\rm (MN)}} &\ =\  \frac{|\Gc|}{B} \Psi_{|\Gc|} 
					\end{align}
				where we recall that $|\Gc|=\frac{K}{B} \gamma+1$. 
			\end{coro}

\begin{figure}[t]\setlength{\abovecaptionskip}{4pt}
			\setlength{\belowcaptionskip}{0pt}
			\centering
			\includegraphics[width=6in]{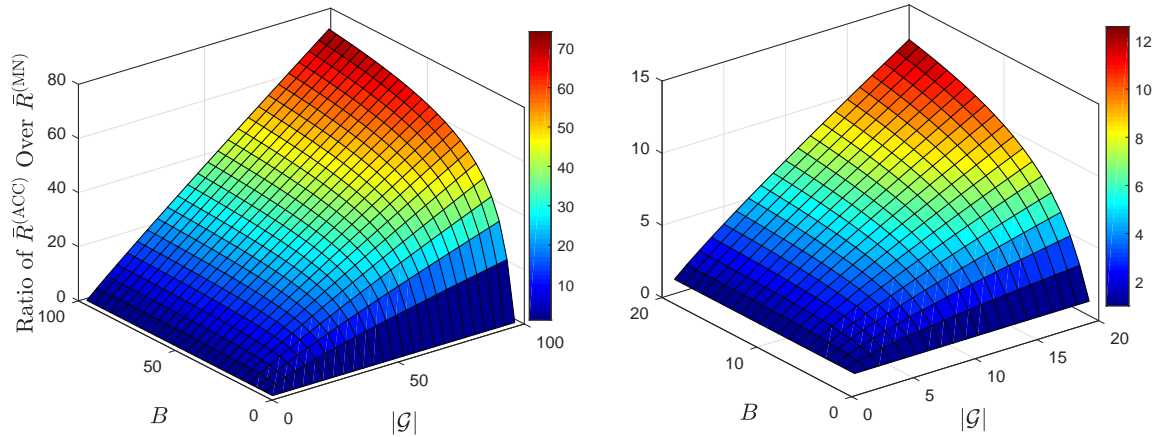}
			\caption{The ACC improvement ($\frac{\Bar{R}^{\rm (ACC)}}{\Bar{R}^{\rm (MN)}} $) over the MN scheme in Cor.~\ref{Delta_coro}, for different $B$ and $|\Gc|$.}\vspace{-0.5cm}\label{Coro1_Gain_Com_fig}
		\end{figure}

			\begin{proof}
				The proof is relegated to Appendix~\ref{app:proof_approx_lowSNR2}.
			\end{proof} 
			The expression in Corollary \ref{Delta_coro} is illustrated in Fig.~\ref{Coro1_Gain_Com_fig} for different values of $B$ and $|\Gc|$.

		\begin{remark}\label{Delta_remark}
		    In Fig.~\ref{Coro1_Gain_Com_fig}, we can see that $\frac{\Bar{R}^{\rm (ACC)}}{\Bar{R}^{\rm (MN)}}$ is concave with respect to $B$, and that this concavity increases with $|\Gc|$. This signals that, for large $|\Gc|$, most of the gain from having $B>1$ is obtained quickly, at relatively small values of $B$. For example, when $|\Gc|=100$ (which is unrealistic), we see that the ACC rate for $B=2$ is up to $20$ times higher than the MN rate ($B=1$).
		\end{remark}		
		
	\subsection{Effective Gain in the Large-$B$ Region}\label{subse:approx_Blarge_Gzero}
	
	    We now move away from the low-SNR regime, and we consider instead the limit of many users. This regime is nicely motivated by the ever increasing density of users in wireless networks. The following shows that, in the limit of many users, the effective gain of the ACC scheme matches -- for any SNR value -- the nominal gain.
	
		\begin{lemma}\label{effectgainTDM}
			For any average SNR~$\rho$, the ACC scheme guarantees 
			\begin{align}\label{eq:delta_approx_largeB}
				\lim_{B\to\infty}\frac{\Bar{R}^{\rm (ACC)}}{\Bar{R}^{\rm (TDM)}}\ 
				= \Lambda\gamma + 1,
			\end{align}			 
			and, thus, its effective gain matches the nominal gain for any value of SNR. 
		\end{lemma}
		\begin{proof}
			The proof is relegated to Appendix~\ref{app:proof_lemma_effectiveGain}.  
		\end{proof} 
	
		We now proceed to compare the ACC scheme with the MN scheme, again in the limit of large~$B$. We will also obtain the low-SNR approximation of this comparison, which nicely captures scenarios such as cell-free or satellite networks, where the majority of the users is distributed in the edge area and/or suffers from heavy path-loss or heavy shadowing.

	    \begin{lemma}\label{codedgain}
			In a setting with $\Lambda$ caches and $K=\Lambda B$ users, and for any average SNR~$\rho$, the ratio $\frac{\Bar{R}^{\rm (ACC)}}{\Bar{R}^{\rm (MN)}}$ satisfies 
    			\begin{align}\label{eq:delta_approx_largeB_a}
    				\lim_{B\to\infty}\frac{\Bar{R}^{\rm (ACC)}}{\Bar{R}^{\rm (MN)}} 
    				    & = \exp\left(\frac{1-|\Gc|}{\rho}\right)
    						\frac{{\rm Ei}\left(-\frac{1}{\rho}\right)}{{\rm Ei}\left(-\frac{|\Gc|}{\rho}\right)}. 
    			\end{align} 
			Furthermore, it holds that 
    			\begin{align}\label{eq:delta_approx_largeB_b}
    				\lim_{\rho\to 0}\ \lim_{B\to\infty}\ \frac{\Bar{R}^{\rm (ACC)}}{\Bar{R}^{\rm (MN)}}
    				    & = \Lambda\gamma + 1.
    			\end{align}			
		\end{lemma}
		\begin{proof}
			The proof is relegated to Appendix~\ref{app:proof_lemma_effectiveGainMN}. Note that~\eqref{eq:delta_approx_largeB_b} follows from~\eqref{eq:delta_approx_largeB_a}, but the same conclusion can be seen directly by combining Proposition~\ref{prop:collapse_caching_gains} and Lemma~\ref{effectgainTDM}. 
		\end{proof} 
		
		\begin{remark}\label{remark:acc_gain}
		    The key for recovering the nominal gain is that a larger $B$ implies a smaller fluctuation around the average transmission rate within a user group, which inherently reduces the impact of the worst-user (or worst-group) bottleneck.
		 \end{remark}

	\subsection{High-Fidelity Approximation of $\bR^\eacc$ for Any SNR Value} \label{subse:approx_largeB}
		The previous subsections offered crisp and insightful approximations of the performance of the ACC scheme. We now take a step back and seek to provide high-accuracy approximations that can be evaluated very easily. 
		
		Indeed, both the exact value of $\bR^\eacc$ in Lemma~\ref{lem:exact} and the approximation at low SNR in Lemma~\ref{lem:ear_approxLow} have time-consuming implementations when $B$ is large. 
		To counter this, we now provide a simple but very precise large-$B$ approximation of $\bR^\eacc$, which accurately approximates the average rate even if $B$ is relatively small.
		This expression involves the well-known Q-function $Q(\cdot)$, i.e.,  the tail  distribution  function  of  the  standard  normal  distribution, and the Meijer's G-function ${\rm G}^{\cdot,\cdot}_{\cdot,\cdot}(\cdot)$ defined in \cite[Eq. (9.301)]{Gradshteyn}. 
		
		Before presenting the new approximation, let us use $H_{|\Gc|}$ to denote the expectation of the maximum of $|\Gc|$ i.i.d. standard normal random variables. 
		Consequently, the expectation of the minimum of such set of variables is given by $-H_{|\Gc|}$. 		
		We can now present our next result. 
		\begin{lemma}\label{lem:approx_expect_rate_BS}
			In the large-$B$ regime, the average rate of the ACC scheme can be approximated by 
				\begin{align}\label{R_min_exact}
					\Bar{R}^{\rm (ACC)} \approx 
					\frac{|\Gc|}{\ln 2} \left( \mu - \frac{\sigma}{\sqrt{B}} \times H_{|\Gc|} \right),
				\end{align}
			where $\mu$ and $\sigma$ respectively represent the average and the standard deviation of $\ln(1+{\rm SNR}_{g,b})$ for $g \in [\Lambda]$ and $b \in [B]$, which are given by
			\begin{align}
				\mu &= - \exp\left(\frac{1}{\rho}\right) \cdot {\rm Ei}\left(-\frac{1}{\rho}\right), \label{eq:mean_prop_bs}\\
				\sigma &=\sqrt{2 \exp\left(\frac{1}{\rho}\right) {\rm G}^{3,0}_{2,3}\left( \frac{1}{\rho} \left|^{1,1}_{0,0,0}\right. \right) -\mu^2}.\label{eq:var_prop_bs}
			\end{align}
		\end{lemma}

		\begin{proof}
		    See Appendix~\ref{app:proof_lem_approx_expect_BS}.
		\end{proof}
		
		The term $H_{|\Gc|}$ is given by the following integral form, 
				\begin{align}\label{H_int}
					H_{|\Gc|} = \frac{-|\Gc|}{\sqrt{2 \pi}} \int_{-\infty}^{+\infty} y \Big(Q(y)\Big)^{|\Gc|-1} \exp\left(-\frac{y^2}{2}\right)  \dd y.
				\end{align}
		The proof of~\eqref{H_int} is relegated to Appendix~\ref{app:proof_lem_approx_expect_BS}. 
		
		At this point, we note that the value of $H_{|\Gc|}$ for $|\Gc|=1,2,3,4,5$ is known and  is given by the following table that can be found in~\cite[Sec. 5.16]{Finch2003}.
			
			\begin{table}[!htbp]
				\centering\caption {Value of $H_{|\Gc|}$ for  $|\Gc|\leq 5$}\label{tbl:table_value}\vspace{-1ex}
				\scalebox{1.2}{
				\begin{tabular}{ c| c c c c c c}  
					\hline         
					$|\Gc|$  & 1 &2 &3 &4 &5 
					\\
					\hline  $H_{|\Gc|}$ &0 
					&$\pi^{-1/2}$  &$\frac{3}{2} \pi^{-1/2}$ 
					&$3 \pi^{-3/2} \cos^{-1}\left(-\frac{1}{3}\right)$ 
					&$\frac{5}{2} \pi^{-3/2} \cos^{-1}\left(-\frac{23}{27}\right)$%
					\\
					\hline     	
				\end{tabular}}\vspace{-0.5cm}
			\end{table}
			
		\noindent For larger values of $|\Gc|$, there are not known closed-form expressions, but it is known (cf.~\cite{kamath2015bounds}) that one can have a simple approximation by substituting $H_{|\Gc|}$ by $\sqrt{2 \ln(|\Gc|)}$. This approximation is based on the fact that $H_{|\Gc|}$ is bounded as $ \frac{1}{\sqrt{\pi \ln 2}} \sqrt{\ln(|\Gc|)} \leq H_{|\Gc|}\leq  \sqrt{2 \ln(|\Gc|)} $, and that~$\lim_{|\Gc|\to\infty}  \frac{H_{|\Gc|}}{\sqrt{\smash[b]{\ln(|\Gc|)}}} = \sqrt{2}$ (cf.~\cite{kamath2015bounds}). 
		
		In order to obtain a better approximation of $H_{|\Gc|}$ than $\sqrt{2 \ln(|\Gc|)}$ -- which is simple but only accurate for large values of~$|\Gc|$ --, a very interesting approximation is to adopt the Gauss-Hermite quadrature (GHQ)~\cite[Ch. 9]{Wikipedia_Gaussian}, which nicely balances high accuracy and low complexity. Applying this method to the specific integral form in~\eqref{H_int} yields 
		\begin{align}
		    H_{|\Gc|} \approx \frac{-\sqrt{2}|\Gc|}{\sqrt{\pi}} \sum\nolimits_{v=1}^{V} \omega_v x_v  \left(Q(\sqrt{2}x_v)\right)^{|\Gc|-1}, \label{eq:GHQ_eq_1}
		\end{align}
		where $V$, $x_v$, and $\omega_v$ are the summation terms, sample points and weights in the GHQ, respectively. Generally speaking, we can get an approximate result with  high accuracy by summing up several terms in the GHQ.

\section{Numerical Results}\label{se:num_analysis}
	In the following, we illustrate through numerical analysis both the exact results and the previously obtained approximations. The derived approximations on the average rate are computationally efficient, can handle large-dimensional problems, and, as we will show via Monte-Carlo simulations, tightly approximate the true performance of the algorithms.
	
	To motivate the values of $B$ that we use, let us consider a scenario with $\gamma=10\%$ and a realistic subpacketization limit of about $10^5$ (for a file size of $10^8$~bytes, this implies an atomic sub-file size of about $1000$~bytes). This gives $\Lambda=\argmax_{x\in\Zb} \big\{\binom{x}{0.1x}< 10^5\big\}\approx 40$, which means that having $K=800$ users reasonably allows for $B$ up to~$20$. Such (or even higher) values of $K$ are motivated by several different scenarios~\cite{Poularakis2016,shafiq2013first}. 
	In order to obtain the simulation results with high accuracy,  $10^6$ channel states are generated and averaged over Rayleigh fading.

    \subsection{Effective Gains With Respect to TDM}
	
\begin{figure}[t]\centering
					\begin{subfigure}[t]{.499\textwidth}\centering							
					\includegraphics[width=3.2 in]{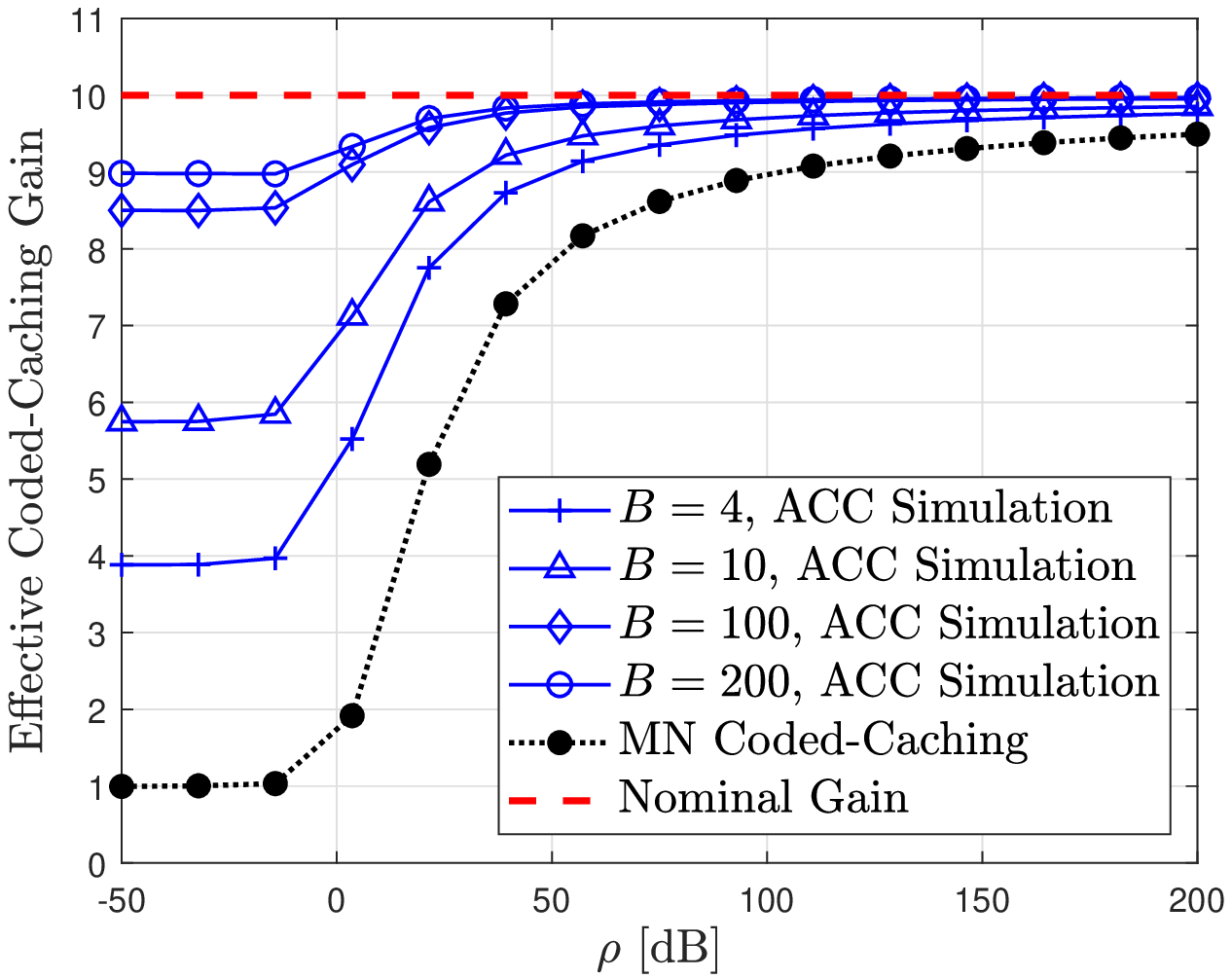}
					\end{subfigure}
					\begin{subfigure}[t]{.499\textwidth}\centering
				        \includegraphics[width=3.2 in]{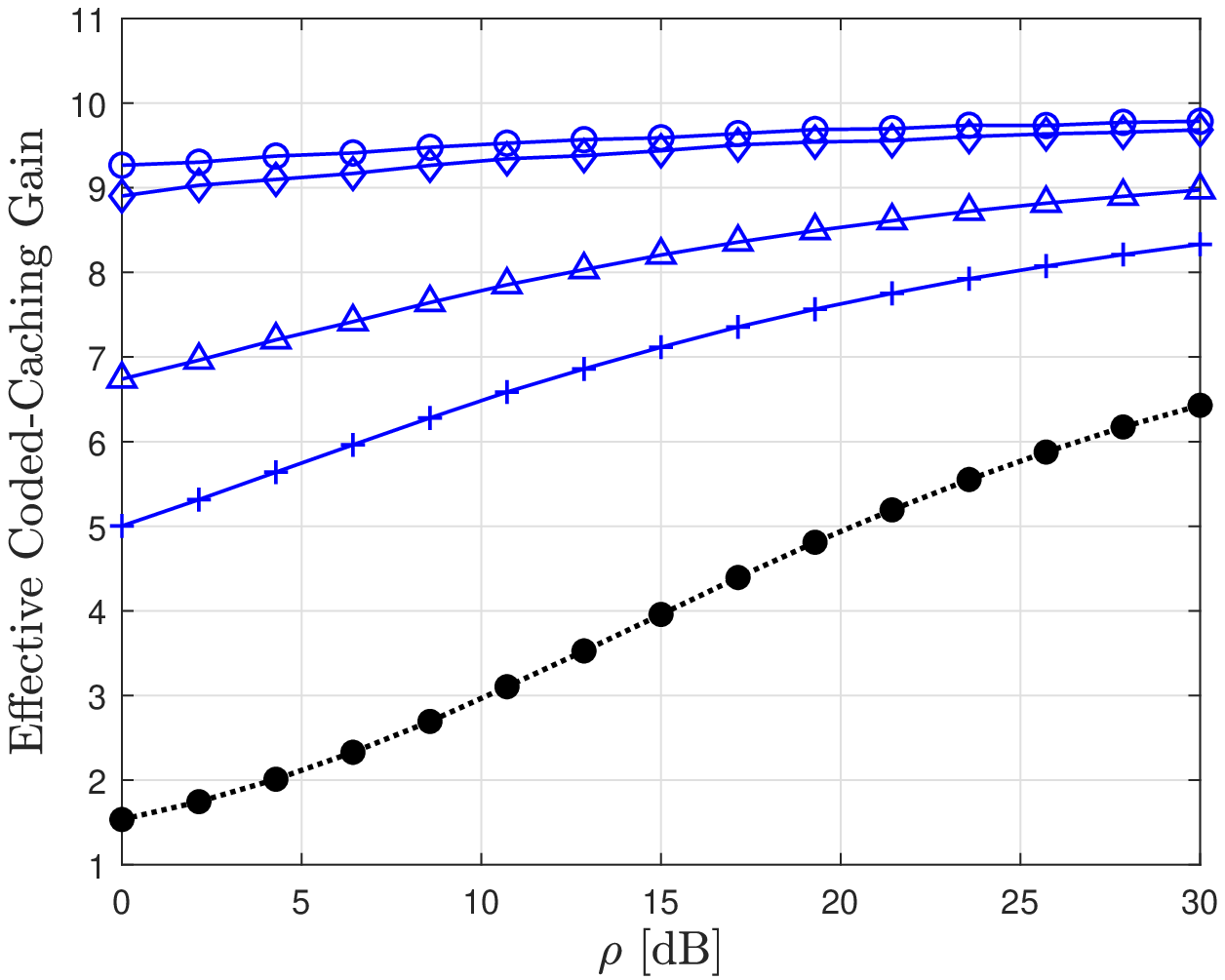}
					\end{subfigure}\vspace{-2ex}
				\caption{Effective gain versus $\rho$ for $|\Gc|=10$. Right-side plot focuses on realistic SNR values.} \vspace{-0.5cm}\label{Gain_MN_fig}
			\end{figure}

\begin{figure}[t]\centering
					\begin{subfigure}[t]{.499\textwidth}\centering		
					\includegraphics[width=3.2 in]{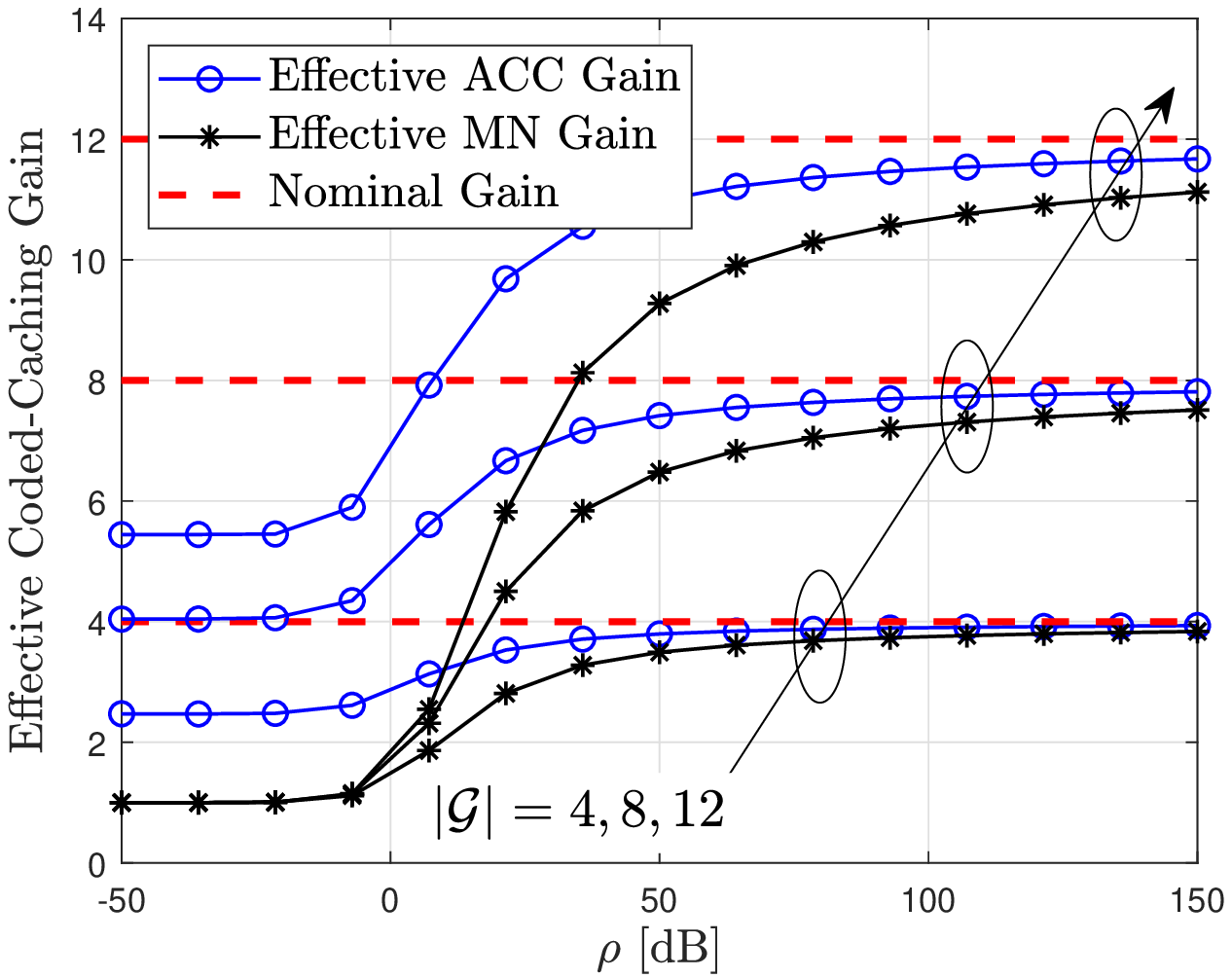}
					\end{subfigure}~%
					\begin{subfigure}[t]{.499\textwidth}\centering					
					\includegraphics[width=3.2 in]{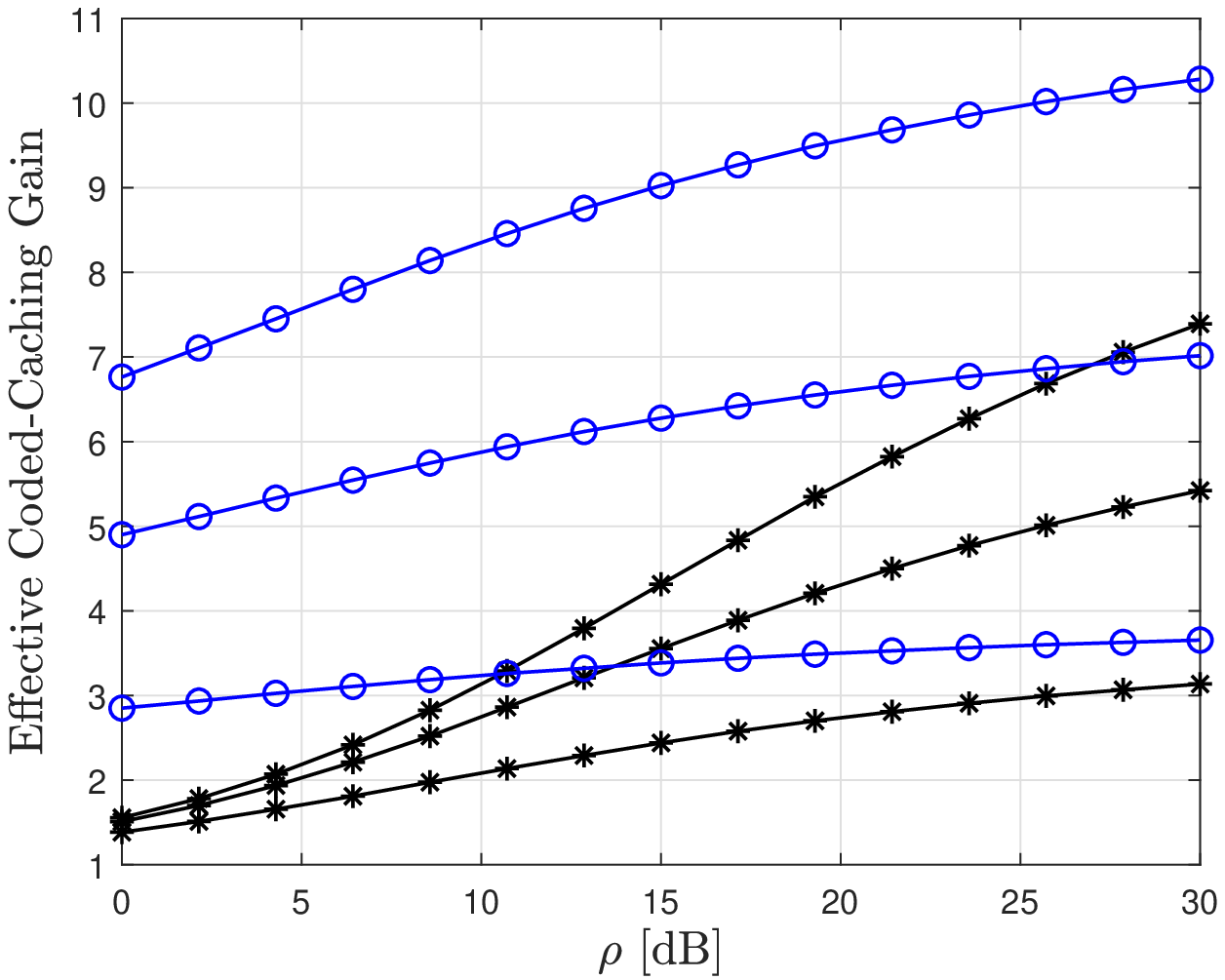}
					\end{subfigure}~\vspace{-1.5ex}
				\caption{Effective gain versus $\rho$ for $B=6$. Right-side plot focuses on realistic SNR values.}\vspace{-0.5cm}\label{Coded_Cache_Gain}
			\end{figure}

    	In Figs. \ref{Gain_MN_fig}--\ref{Coded_Cache_Gain}, we present the effective coded-caching gains of the ACC and MN schemes versus $\rho$, for different values of $B$ and different nominal gains ($|\Gc|$). 
    	As expected, the effective gains of both the ACC scheme and the MN scheme converge to the nominal gain as $\rho$ increases. 
    	However, the convergence of the ACC scheme is much faster than that of the MN scheme and, furthermore, the convergence of the ACC scheme becomes faster as $B$ grows.  
    	
    	From the same figures, it is also worth noting that, when $\rho$ is relatively small, the effective coded-caching gains of both schemes arrive to a flat lower bound. 
    	The lower bound for the ACC scheme is notably greater and improves as either $B$ or $|\Gc|$ become bigger.  
	    However, this behavior does not extend to the MN scheme, which is consistent with the result of Proposition~\ref{prop:collapse_caching_gains} which states that the effective gain of the MN scheme collapses at low SNR, regardless of the value of  the high-SNR caching gain $|\Gc|$. 
	    
	    Moreover, in Fig.~\ref{Coded_Cache_Gain}, we can see that for the MN scheme the worst-user effect is amplified as $|\Gc|$ increases. Then, Figs.~\ref{Gain_MN_fig}--\ref{Coded_Cache_Gain} show that the advantages of the ACC scheme in terms of average rate are still significant even for a small group size ($B=4,6$).

    \subsection{Approximations on the Average Rate $\Bar{R}^\eacc$}

\begin{figure}[t]\centering
					\begin{minipage}[t]{.5\textwidth}\centering
						\includegraphics[width= 3.2 in]{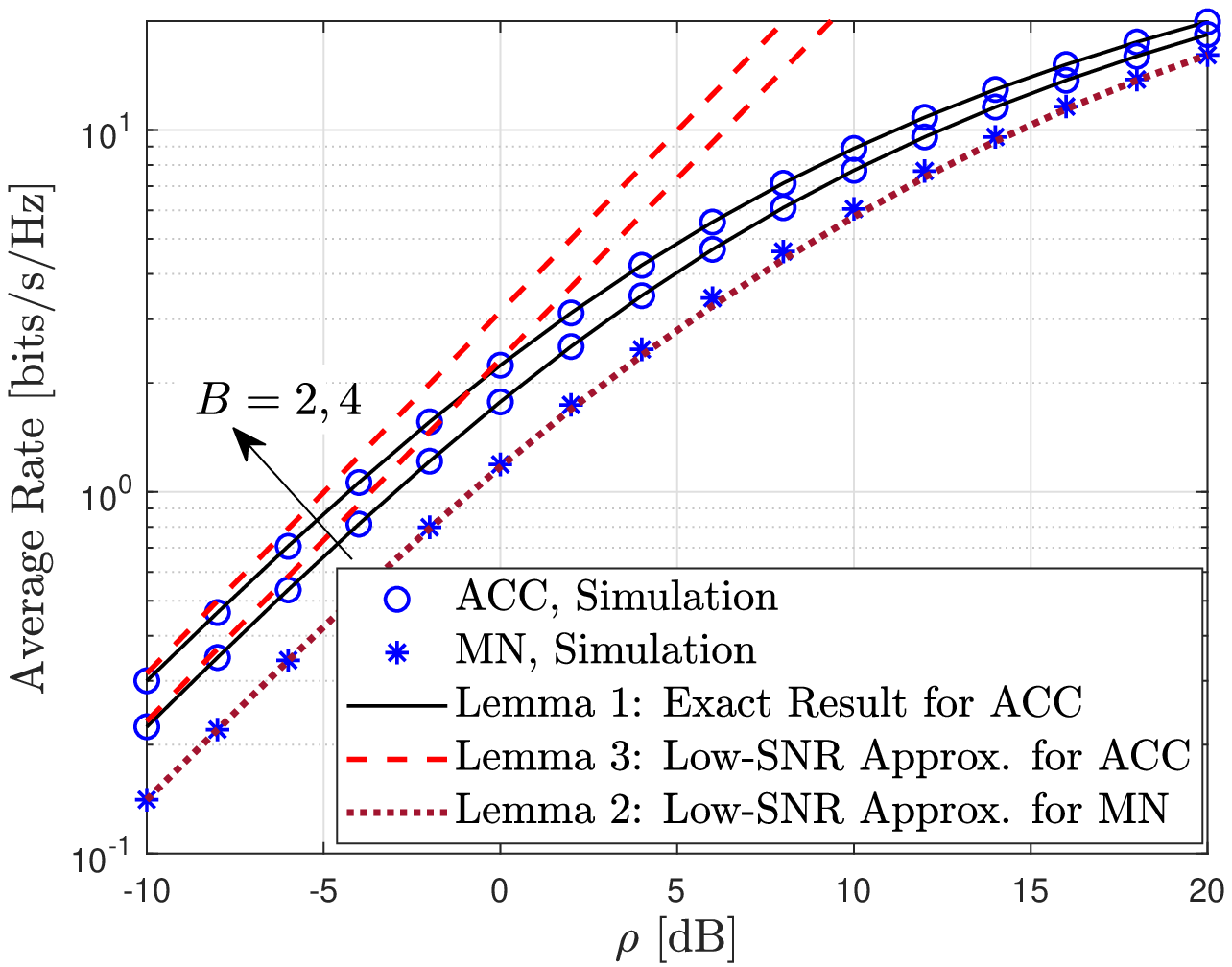}\vspace{-2ex}
						\captionof{figure}{$\racc$  versus $\rho$ for $|\Gc|=4$.}\label{EAR_fig_com} 
					\end{minipage}%
					\begin{minipage}[t]{.5\textwidth}\centering
						\includegraphics[width= 3.2 in]{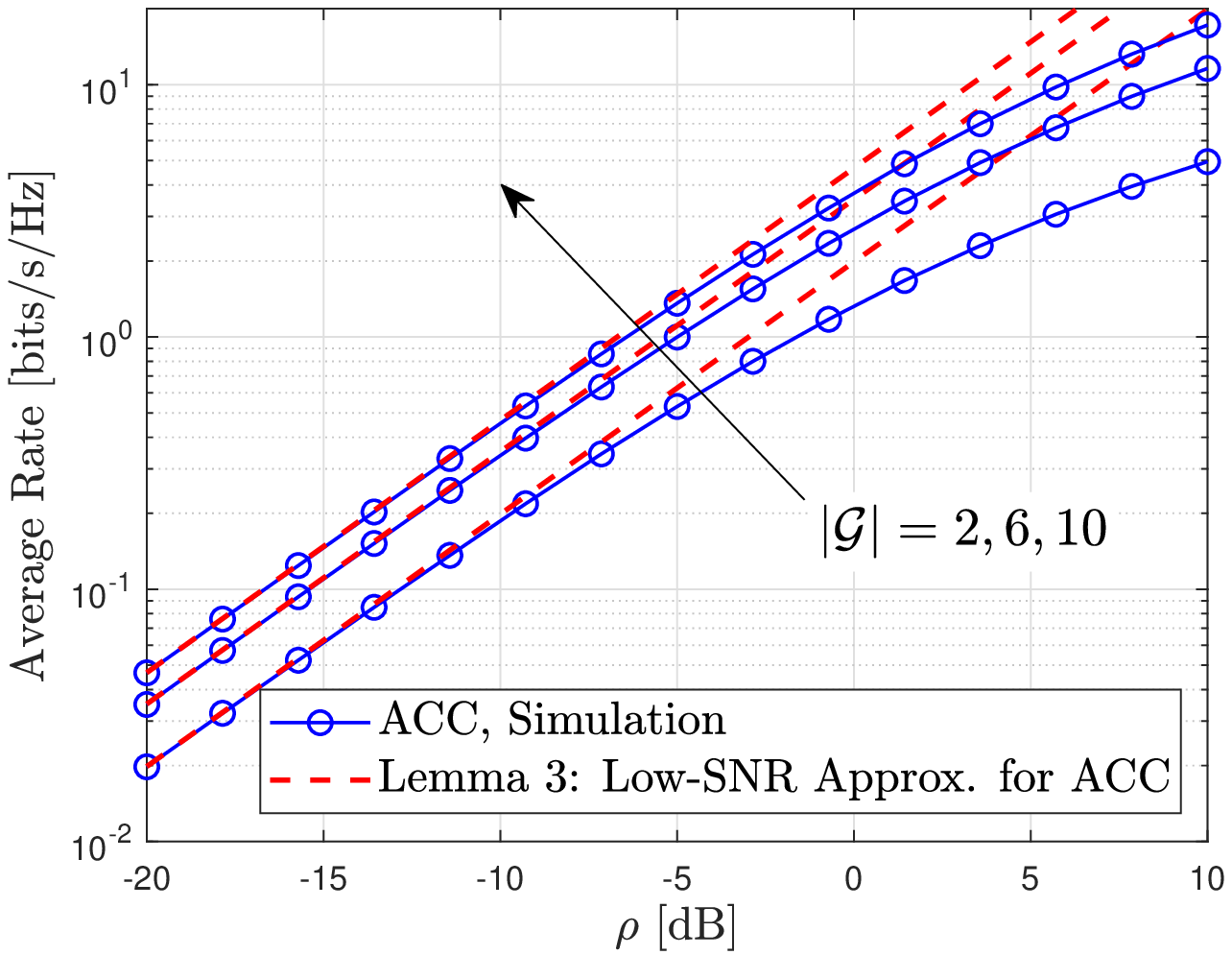}\vspace{-2ex}
						\captionof{figure}{$\racc$  versus $\rho$ for $B=3$.}\label{EAR_fig_G}
					\end{minipage}\vspace{-.9cm}
				\end{figure}

\begin{figure}[t]\centering
					\begin{minipage}[t]{.5\textwidth}\centering
						\includegraphics[width= 3.2 in]{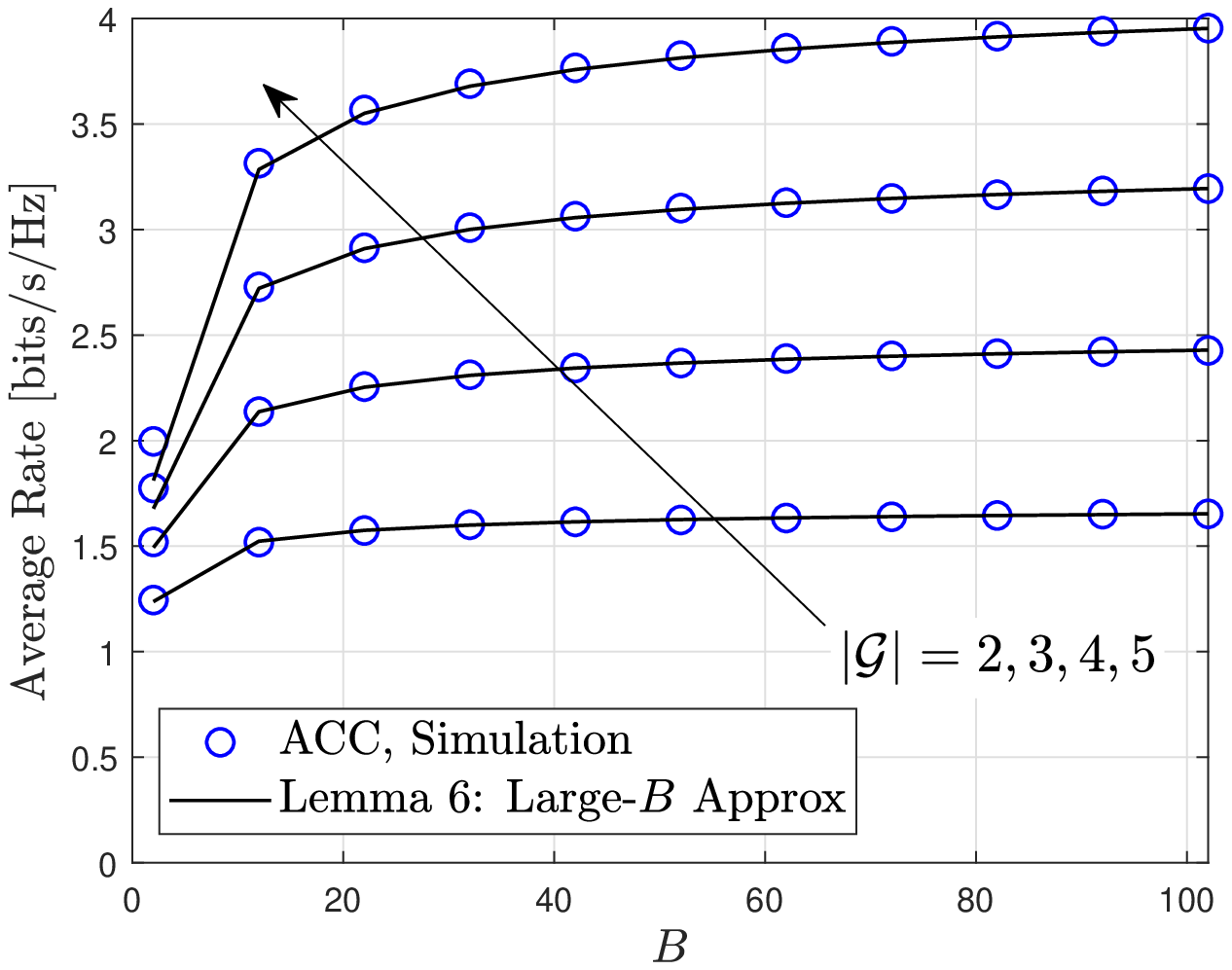}\vspace{-2ex}
						\captionof{figure}{$\racc$ versus $B$ for $\rho=0$ dB.}\label{EAR_fig_BG}
					\end{minipage}%
					\begin{minipage}[t]{.5\textwidth}\centering
						\includegraphics[width= 3.2 in]{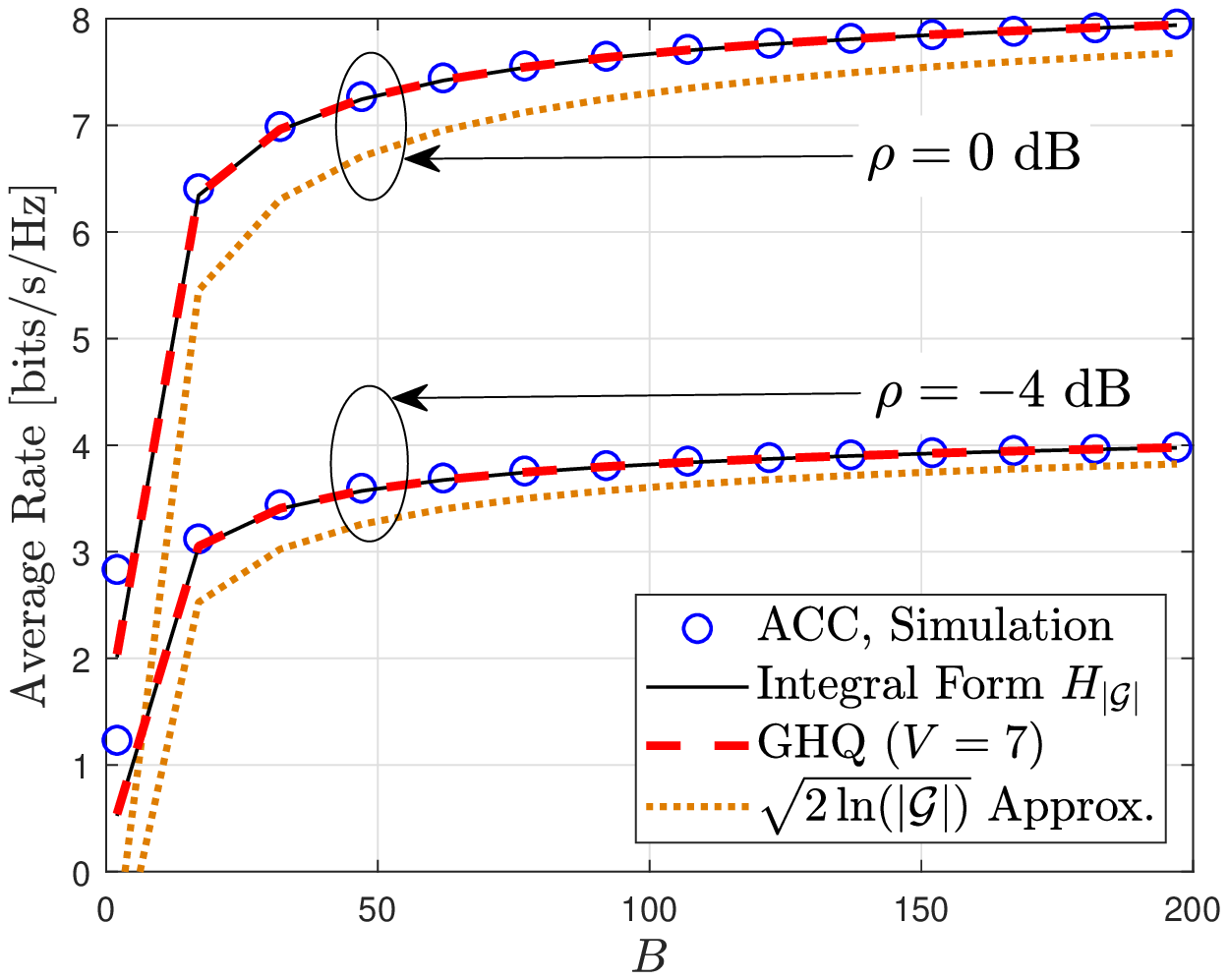}\vspace{-2ex}
						\captionof{figure}{$\racc$ versus $B$ for $|\Gc|=10$.}\label{EAR_fig_gamma}
					\end{minipage}\vspace{-0.5cm}
				\end{figure}

			In Figs. \ref{EAR_fig_com}--\ref{EAR_fig_gamma}, we validate the derived analytical approximations and highlight some interesting trends and comparisons. 
			First, Fig. \ref{EAR_fig_com} shows the average rate $\Bar{R}^\eacc$ versus $\rho$ for different values of $B$. Note that, for $B=1$, $\Bar{R}^\eacc=\Bar{R}^\emn$. 
			For comparison, we also plot in  Fig.~\ref{EAR_fig_com} the different derived approximations. 
			Specifically, Fig.~\ref{EAR_fig_com} displays the simulated result (circle and asterisk symbols), the exact derived average rate $\Bar{R}^\eacc$ in Lemma~\ref{lem:exact} (solid line), the low-SNR multinomial approximation in Lemma~\ref{lem:ear_approxLow} (dashed line), and the low-SNR second-order approximation for $\Bar{R}^\emn$ in Lemma~\ref{lem:ear_mn_approxLow} (dotted line). 
			The rate enhancement due to the ACC scheme is exhibited by comparing the results of Lemma~\ref{lem:exact} and Lemma~\ref{lem:ear_mn_approxLow} (solid and dotted lines, respectively). 
			Fig.~\ref{EAR_fig_com} shows that the accuracy of the approximation for $\Bar{R}^\emn$ in Lemma~\ref{lem:ear_mn_approxLow} is better than the approximation for $\Bar{R}^\eacc$ in Lemma~\ref{lem:ear_approxLow}, mainly because Lemma~\ref{lem:ear_approxLow} considers a first-order approximation. 
			Fig. \ref{EAR_fig_G} reveals that the approximation derived in Lemma~\ref{lem:ear_approxLow} becomes more accurate as $|\Gc|$ increases, which indicates that the value of $\rho$ at which the nonlinear part of the average rate becomes significant increases as $|\Gc|$ increases.

			The large-$B$ approximation of $\Bar{R}^\eacc$ from Lemma~\ref{lem:approx_expect_rate_BS} is validated in Fig.~\ref{EAR_fig_BG}, where the average rate is plotted for different $|\Gc|$\footnote{The values of $H_{|\Gc|}$ for $|\Gc|=2,3,4,5$ are taken from Table~\ref{tbl:table_value}.}. 
			This large-$B$ approximation tightly approximates the simulation results, even for a small $B$. In fact, this approximation is extremely tight for any value of B bigger than~$1$. 
			To further demonstrate the accuracy of Lemma~\ref{lem:approx_expect_rate_BS}, we show in Fig.~\ref{EAR_fig_gamma}  the results derived by using $i)$ the integral calculation in \eqref{H_int}, 
			$ii)$ the GHQ method in~\eqref{eq:GHQ_eq_1}, 
			and $iii)$ the $\sqrt{2 \ln (|\Gc|)}$ approximation of $H_{|\Gc|}$ for $|\Gc|>5$. 

\begin{figure}[t]\centering
					\begin{minipage}[t]{.5\textwidth}\centering
						\includegraphics[width= 3.2 in]{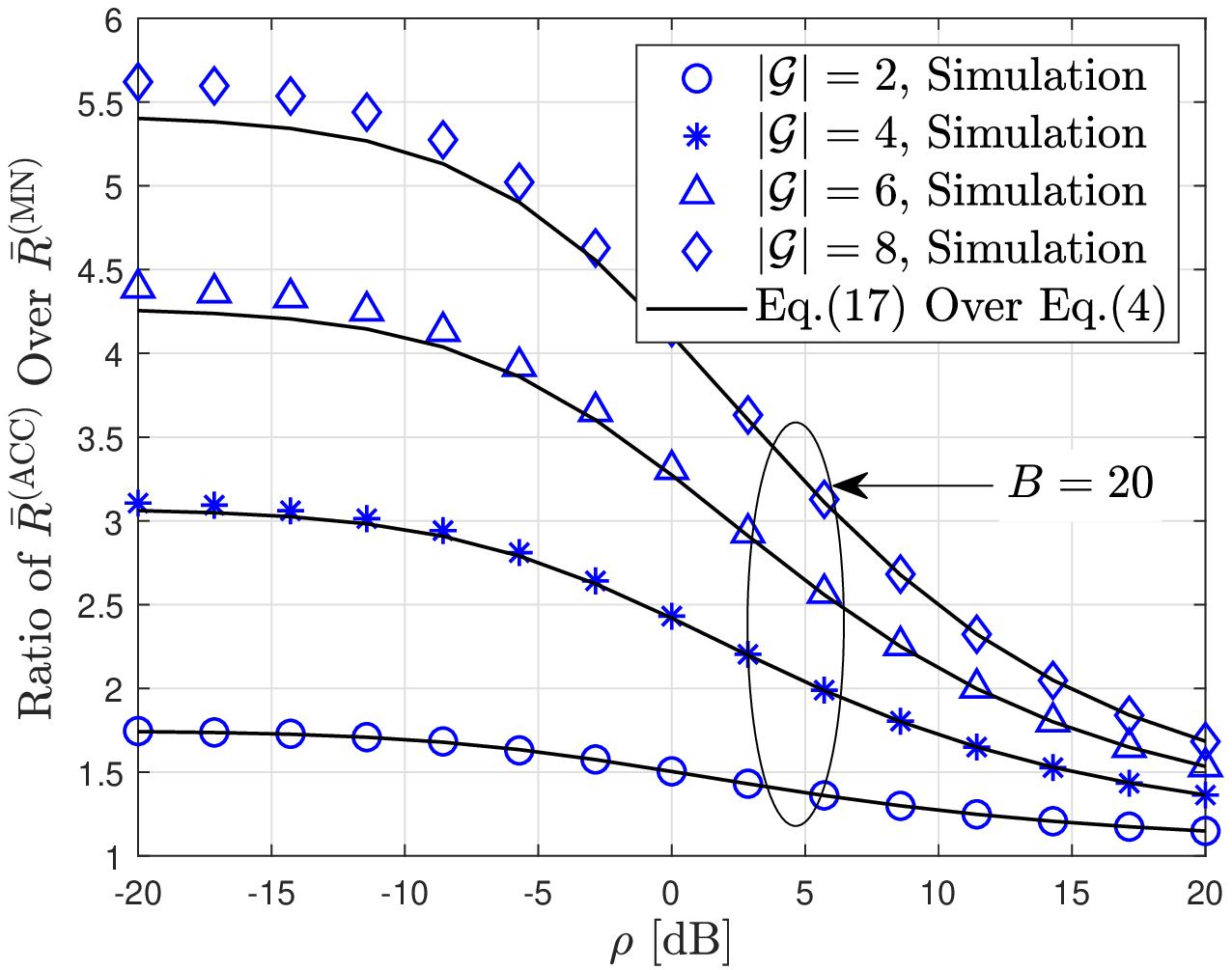}\vspace{-2ex}
						\captionof{figure}{$\frac{\Bar{R}^{\rm (ACC)}}{\Bar{R}^{\rm (MN)}}$ versus $\rho$ for $V=7$ in GHQ.}\label{EAR_G_Com_fig}
					\end{minipage}%
					\begin{minipage}[t]{.5\textwidth}\centering
						\includegraphics[width= 3.2 in]{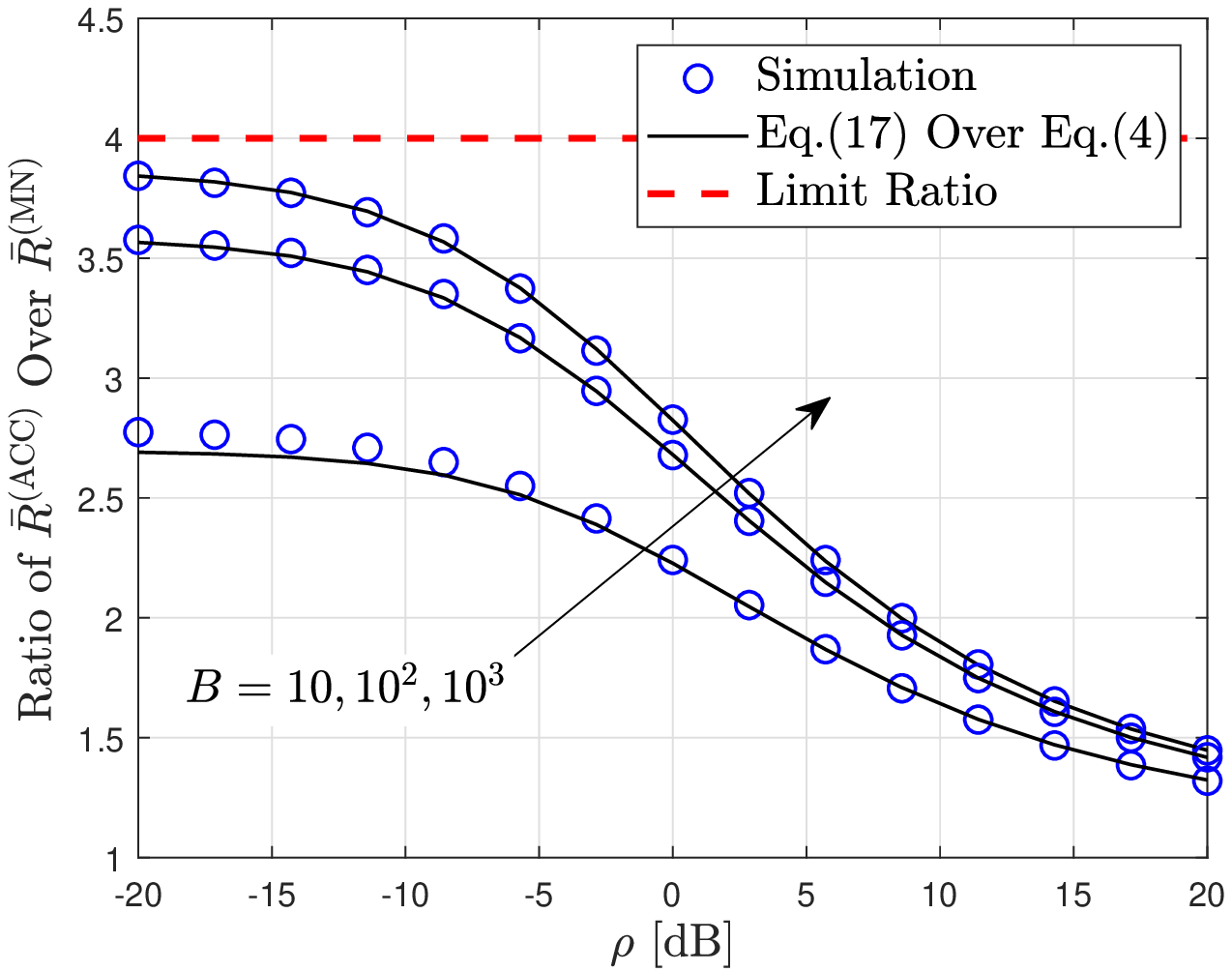}\vspace{-2ex}
						\captionof{figure}{$\frac{\Bar{R}^{\rm (ACC)}}{\Bar{R}^{\rm (MN)}}$ versus $\rho$ for $|\Gc|=4$.}\label{Limit_Ratio_fig}
					\end{minipage}\vspace{-0.75cm}
				\end{figure}

			After verifying the high accuracy of the approximation in Lemma~\ref{lem:approx_expect_rate_BS}, we exploit it to present some interesting comparisons between the ACC scheme and the MN scheme in Figs.~\ref{EAR_G_Com_fig}--\ref{Limit_Ratio_fig}. 
			In Fig.~\ref{EAR_G_Com_fig}, we can see through the ratio $\frac{\Bar{R}^{\rm (ACC)}}{\Bar{R}^{\rm (MN)}}$ that $\Bar{R}^{\rm (ACC)}$ provides significant boost for realistic SNR values.  
			In order to illustrate  the extent to which this ratio approaches the theoretical gain in the low-SNR regime, we show in Fig.~\ref{Limit_Ratio_fig} the different ratios/improvements achieved by varying~$B$. 

	\section{Conclusions}\label{se:conclusions}
		This work is motivated by the fact that any attempt to successfully adopt wireless coded caching in large-scale settings must account for the effects of low-to-moderate SNR fading channels. Toward this, we first revealed that dedicated caches and XOR-based transmissions may no longer be suitable for various realistic SNR regimes. As we have seen, as the SNR becomes smaller, the effective gains of the MN scheme collapse, irrespective of either the nominal gain or~$K$.  
		We have then proposed a novel and simple scheme that recovers a big fraction of the lost gains and does so for any SNR value. These gains are fully recovered in the regime of many users, again for any SNR value, thus essentially resolving the worst-user bottleneck. The scheme builds on the idea of shared caches, which is an inevitable feature of practical coded caching settings due to the file-size constraint. 
		The gains appear in practical values of SNR and for realistically many users. As seen in Fig.~\ref{Coro1_Gain_Com_fig}, having as few as $B=2$ users per group allows the ACC scheme to approximately double the coded caching gain. As stated before, these gains do not involve user selection, and the corresponding user-grouping is done prior to cache-placement and is oblivious to the demands and of course oblivious to the channel. 
		Finally, the derived expressions are simple but very precise. For example, the low SNR approximation for the MN scheme in Lemma \ref{lem:ear_mn_approxLow} is essentially identical to the actual performance even for SNR values as high as~$20$~dB. Similarly, as Fig.~\ref{EAR_fig_BG} shows, the large-$B$ approximation is almost exact even for values of $B$ as low as~$10$.
				
		In the end, the ACC scheme applies toward showing that properly designed coded caching has the ability to substantially speed up delivery of multimedia content even in the challenging environment of low-to-moderate SNR fading channels.


\appendices
\renewcommand{\thesectiondis}[2]{\Roman{section}:}

\section{Capacity Region of Proposition~\ref{prop:cap_reg_bc}}\label{app:capacity_region}
	The appendix is meant to orient the reader as to how the existing results in~\cite{Tuncel2006} on multicasting with side information\footnote{Several works have considered this Gaussian setting after~\cite{Tuncel2006}. In~\cite{Kramer2007_ITW}, the capacity region was derived for the 2-user case, the 3-user case was studied in~\cite[Group 8, case $\Gc_{18}\bigcup\Gc_{28}$]{Asadi2015_TIT}, and the converse of Prop.~\ref{prop:cap_reg_bc} can be also found in~\cite[Thm. 4]{Yoo2009}. 
	} can be applied to our setting. Using the notation of~\cite{Tuncel2006}, we recover Proposition~\ref{prop:cap_reg_bc} from ~\cite[Thm. 6]{Tuncel2006} by choosing $X^n$ to be $(X_1,\; X_2,\;\cdots,\; X_t)$, $m=n$, setting the side information $Y_i$ to be $Y_i = \{X_\ell\}_{\ell\in[t]\backslash i}$, and applying invertible mappings between $X^n_i$ and $W'_i$ for any $i\in t$. 
	From the maximum entropy theorem~\cite[Thm. 9.6.5]{Cover2006}, we obtain Proposition~\ref{prop:cap_reg_bc}. 
	
	For the achievability part, we proceed as in~\cite{Tuncel2006} and  consider a codebook of $2^{n(\sum_{\ell=1}^{t} R_\ell)}$ codewords. The codewords are denoted by $x^n(w_1,w_2,\cdots,w_L)$, with $w_\ell \in [2^{nR_\ell}]$, for any $\ell\in[L]$. The  letters of the codewords, denoted by $x_j(w_1,w_2,\cdots,w_L)$, $j\in[n]$,  are i.i.d. distributed as $\Nc(0,P)$. 
	Each user can decode its intended message from the received signal and from the (cached) side information using typical set decoding. The intuition behind the successful decoding at a certain user~$i$ is that, after receiving one of the $2^{n(\sum_{\ell=1}^{L} R_\ell)}$ codewords and thanks to the cached information, user~$i$ applies typical decoding over only $2^{nR_i}$ possible codewords.

\section{Proof of Lemma~\ref{lem:exact}}\label{app:proof_lemma_exact}
	Let us start by defining $S_g \triangleq \sum_{b=1}^{B} \ln(1+{\rm SNR}_{g,b})$ for any group~$g\in[\Lambda]$ of users. Also note that we can write the average rate of the ACC scheme as
		\begin{align}\label{eq:rate_bs_minS}
			\Bar{R}^{\rm (ACC)}=\frac{|\Gc|}{B \ln2} \E_{H}\left\{\min_{g \in \mathcal{G}} \{S_g\}\right\}. 
		\end{align}
	For $t \in (-\infty, +\infty)$, the characteristic function (CF) in probability~\cite[Ch. 5]{Papoulis} of $S_g$ is defined~as
		\begin{align}\label{CF_Xg_Def}
			{\rm CF}_{S_g}(t)&=\E \left\{ \exp(\jmath t S_g) \right\} 
				=\E \Big\{ \exp\Big( \jmath t \sum_{b=1}^{B} \ln(1+{\rm SNR}_{g,b}) \Big)   \Big\}  
				= \left[ \E\left\{ (1+{\rm SNR}_{g,b})^{\jmath t}   \right\}     \right]^B\!.
		\end{align}
	Substituting the PDF of ${\rm SNR}_{g,b}$ into \eqref{CF_Xg_Def} yields
		\begin{align}
			{\rm CF}_{S_g}(t)&=\frac{1}{\rho^B}\left[ \int_0^\infty (1+x)^{\jmath t} \exp\left(-\frac{x}{\rho}\right) \dd x   \right]^B 
			\mathop  =  \limits^{(a)} \frac{1}{\rho^B} \exp\left(\frac{B}{\rho}\right) {\rm E}_{-\jmath t}^B \left( \frac{1}{\rho}  \right),
		\end{align}
	where $(a)$ follows from~\cite[Eq. (3.382.4)]{Gradshteyn}. 
	By considering the Gil-Pelaez Theorem~\cite{Pelaez}, the CDF of $S_g$ is obtained as
		\begin{align}
			F_{S_g}(x)= \frac{1}{2}-\frac{1}{\pi} \int_0^\infty \frac{{\rm Im} \left\{ \exp({-\jmath x t}) \frac{\exp({B / \rho})}{\rho^B} {\rm E}_{-\jmath t}^B \left(\frac{1}{\rho}\right) \right\}}{t} \; \dd t.
		\end{align}
      Define $J\triangleq \min_{g \in \mathcal{G}} \{S_g\} =\min_{g \in \mathcal{G}} \big\{\sum_{b=1}^{B} \ln(1+{\rm SNR}_{g,b}) \big\}$. The CDF of $J$ can be expressed by
		\begin{align}\label{CDF_Y_Exact}
			F_J(y) &= \Pr \Big\{ \min_{g \in \mathcal{G}} \{S_g\} \le y \Big\}=1- \Pr \Big\{ \min_{g \in \mathcal{G}} \{S_g\} > y \Big\} 
			=1- \left( \Pr\left\{  S_g>y  \right\}   \right)^{|\Gc|} \notag\\
			&= 1- \left( \frac{1}{2} +\frac{1}{\pi} \int_0^\infty \frac{{\rm Im} \left\{ \exp({-\jmath x t}) \frac{\exp({B / \rho})}{\rho^B} {\rm E}_{-\jmath t}^B \left(\frac{1}{\rho}\right) \right\}}{t} \dd t \right)^{\!\!|\Gc|}\!\!.
		\end{align}
	As $J$ is a non-negative random variable, it holds that $\E\left\{ J \right\} = \E\big\{ {\int_0^J  {\dd x} } \big\} $, and furthermore,
		\begin{align}\label{E_Y}
			\E\left\{ {\int_0^J  {\dd x} } \right\} 
			= \E\left\{ {\int_0^\infty  {\mathbb{I}\left\{ {x \le J} \right\}\dd x} } \right\} 
			= \int_0^\infty  {\E\left\{ {\mathbb{I}\left\{ {x \le J} \right\}} \right\}\dd x}  
			= \int_0^\infty  {\left[ {1 - {F_J}\left( y \right)} \right]\dd y},
		\end{align}
	where $\mathbb{I}\{\cdot\}$ denotes the indicator function, which, for  claim $\mathcal{A}$, takes the value
		$\mathbb{I}\{\mathcal{A}\}=1$ if $\mathcal{A}$ is true and 	$\mathbb{I}\{\mathcal{A}\}=0$ otherwise. 
	Combining \eqref{CDF_Y_Exact} and \eqref{E_Y} yields that the expectation of $J$ is given by 
		\begin{align}
			\E\{J\}= 
			 \int\nolimits_0^\infty \left( \frac{1}{2} +\frac{1}{\pi} \int\nolimits_0^\infty \frac{{\rm Im} \left\{ \exp({-\jmath x t}) \frac{\exp({B / \rho})}{\rho^B} {\rm E}_{-\jmath t}^B \left(\frac{1}{\rho}\right) \right\}}{t} \dd t \right)^{\!\!|\Gc|}\! \dd y.
		\end{align}
	It follows from~\eqref{eq:rate_bs_minS} that $\Bar{R}^{\rm (ACC)}=\frac{|\Gc|}{B \ln2} \E\{J\}$, which gives~\eqref{AER_BS_Int} by considering the integral form of $\E\{J\}$, and therefore Lemma~\ref{lem:exact} is proven. \qed

\section{Proofs for Section~\ref{se:low_snr} and Section~\ref{subse:approx_Blarge_Gzero}}\label{app:proof_general}

    \subsection{Proof of Lemma~\ref{lem:ear_mn_approxLow}}\label{app:ear_mn_approxLow}
    	The fact that~$\SNR_g$ is distributed as $\Exp(\nicefrac{|\Gc|}{\rho})$ implies that ${\rm Var}(\min_{g \in \mathcal{G}}\{\SNR_g\})=\nicefrac{\rho^2}{|\Gc|^2} =o(\rho)$. Thus, in a similar way as in~\cite[Eq. (4)]{Zhao_CL_GK}, in the low-SNR region we can  approximate $\rmn$ by its robust approximation based on the Taylor series: Let $P(X)$ be a real-valued function with respect to a random variable $X$ with mean $\mu_X$ and variance $\sigma_X^2$. The expectation of $P(X)$ can be tightly approximated in the low $\sigma^2_X$ region as 
    		\begin{align}\label{eq:approx_low_var}
    			\E\{P(X)\} \approx P(\mu_X)+\frac{\sigma_X^2}{2} \left. \frac{\partial^2 P(X)}{\partial X^2} \right|_{X=\mu_X}
    		\end{align}
    	where  $\frac{\partial^2 P(X)}{\partial X^2}$ stands for the second derivative of $P(X)$ with respect to $X$ (cf.~\cite{Holtzman}).
    
    	Consider that $P(X)=\frac{|\Gc|}{\ln 2} \ln \left(1+ \min_{g \in \mathcal{G}} \left\{{\rm SNR}_{g}\right\}\right)$ and $X=\min_{g \in \mathcal{G}}\{{\rm SNR}_g\}$. By adopting the robust approximation in~\eqref{eq:approx_low_var}, $\Bar{R}^{\rm (MN)}$ can be tightly approximated at low SNR by \eqref{EAR_Robustapprox}. \qed
	
    \subsection{Proof of Proposition~\ref{prop:collapse_caching_gains}}\label{app:gain_mn_approxLow}
	     Given that $-e^{-x}\ln(1+\frac{1}{x})< {\rm Ei}(-x) < \frac{-e^{-x}}{2}\ln(1+\frac{2}{x})$\cite{abramowitz55handbook}, we can upper bound the numerator and lower bound the denominator of the exact expression of  $\frac{\bar R^{\rm (MN)}}{\bar R^{\rm (TDM)}}$ in~\eqref{Omega_MN_exact} to obtain that 
				\begin{align}\label{EAR_TDM_Intro3}
			    \lim_{\rho \to 0} \frac{\bar R^{\rm (MN)}}{\bar R^{\rm (TDM)}}
			    \le \lim_{\rho \to 0} \frac{|\Gc|}{2} \frac{\ln\left(1+\frac{2\rho}{|\Gc|} \right)}{\ln (1+\rho)}=1.
				\end{align}								
			By interchanging the bounds to lower bound the ratio, we obtain that the limit is also lower bounded by $1$, which concludes the proof of Proposition~\ref{prop:collapse_caching_gains}. \qed

    \subsection{Proof of Lemma~\ref{lem:ear_approxLow} }\label{app:proof_approx_lowSNR1}
       From the definition of $S_g$ in Appendix~\ref{app:proof_lemma_exact}, we see that
        \begin{align}
        	  \E\{S_g\} &= \E\big\{\sum\nolimits_{b=1}^B {\rm SNR}_{g,b}\big\} + \E\big\{\sum\nolimits_{b=1}^B \big(\ln(1+\SNR_{g,b})-{\rm SNR}_{g,b}\big)\big\} \\
        	      & = \E\big\{\sum\nolimits_{b=1}^B {\rm SNR}_{g,b}\big\} + o(\rho). \label{eq:approx_exp}
    		\end{align}
    	In the above, ~\eqref{eq:approx_exp} is obtained from Lebesgue's Dominated Convergence Theorem\cite[Thm.~16.4]{Billingsley1995} as follows: First, we know that $\lim_{x\to 0}(\ln(1+x) - x)/x = 0$, and hence $\ln(1+x) - x = o(x)$ as $x \to 0$. In order to prove that the expectation is also~$o(\rho)$ as $\rho \to 0$, we need to prove that $|\ln(1+x) - x|$ is bounded by some integrable function. For that, we first note that $\ln(1+x) - x\leq 0$ for any $x>0$.
    	Thus, it follows that  $|\ln(1+\SNR_{g,b}) - \SNR_{g,b}| \leq |\SNR_{g,b}|$, which satisfies that $\E\{|\SNR_{g,b}|\} = \rho<\infty$. Hence, we can apply the Dominated Convergence Theorem and obtain that $ \E\big\{\sum\nolimits_{b=1}^B \big(\ln(1+\SNR_{g,b})-{\rm SNR}_{g,b}\big)\big\} =o(\rho)$. 
    	
     	Since ${\rm SNR}_{g,b}$ follows a distribution ${\rm Exp}(\frac{1}{\rho})$, then the sum $\sum\nolimits_{b=1}^B {\rm SNR}_{g,b}$ follows a ${\rm Gamma}(B, \rho)$ distribution, with shape and scale parameters $B$ and $\rho$ respectively.
    	Then the CDF of $\Phi\triangleq \min_{g \in \mathcal{G}} \{\sum\nolimits_{b=1}^B {\rm SNR}_{g,b} \}$ takes the form
    		\begin{align} 
    			F_\Phi(y) & = 1- \left(\frac{1}{\Gamma(B)}  \Gamma\left(B, \frac{y}{\rho}\right)  \right)^{|\Gc|} 
    			\mathop  =  \limits^{(a)} 1-\left( \exp\left(-\frac{y}{\rho}\right)  \sum\nolimits_{t=0}^{B-1} \frac{y^t}{t! \,\rho^t} \right)^{|\Gc|},
    		\end{align}
    	where $\Gamma(\cdot,\cdot)$ denotes the upper incomplete Gamma function \cite{Gradshteyn},  and $(a)$ follows from~\cite[Eq. (8.352.2)]{Gradshteyn} since $B$ is a positive integer. 
    	For  $\bb\in\Zb^{B}$, let $b_t \triangleq \bb(t)\geq 0$, $t\in[B]$, denote its $t$-th element. Recalling that ${n \choose {\bb}}\triangleq \frac{n!}{b_1!b_2!\cdots b_B!}$, we apply the  Multinomial theorem \cite{Kataria} to get
    		\begin{align} 
    			&F_\Phi(y) 
    			=  1- \exp \left(-\frac{|\Gc|y}{\rho}\right) 
    			\sum\nolimits_{||{\bf b}||_1=|\Gc|} \! {|\Gc| \choose {\bf b}} \frac{\rho^{-\sum_{t=1}^{B} (t-1) b_t}}{\prod_{t=1}^{B} ((t-1)!)^{b_t}} y^{\sum_{t=1}^{B} (t-1) b_t}.
    		\end{align}
			In view of the relationship between the CDF and the expectation in \eqref{E_Y}, 
			the average rate of the ACC scheme can be approximated in the low-SNR region by
    		\begin{align} 
    			\Bar{R}^{\rm (ACC)} &= \frac{|\Gc|}{B \ln2} \big(\E\{\Phi\}+o(\rho)\big) =\frac{|\Gc|}{B \ln2} \int_0^\infty \big[1-F_\Phi(y)\big] \dd y +o(\rho)\notag\\
    			&= \frac{|\Gc|}{B \ln2} \sum_{||{\bf b}||_1=|\Gc|}\! {|\Gc| \choose {\bf b}} \frac{\rho^{-\sum_{t=1}^{B} (t-1) b_t}}{\prod_{t=1}^{B} ((t-1)!)^{b_t}} \int_0^\infty\!\! \exp \left(-\frac{|\Gc|y}{\rho}\right)  y^{\sum_{t=1}^{B} (t-1) b_t} \dd y + o(\rho), \label{eq:apprx_rate_lowCDF}
    		\end{align}
    	which can be solved by using the definition of Gamma function~\cite[Eq. (8.312.2)]{Gradshteyn}. \qed

    \subsection{Proof of Corollary~\ref{Delta_coro} }\label{app:proof_approx_lowSNR2}
      From Lemma~\ref{lem:ear_approxLow} we have that $\Bar{R}^\eacc = \frac{\rho |\Gc|}{B \ln2}\, \Psi_{|\Gc|} + o(\rho)$ and also that $\bar R^{\rm (TDM)}=\bar R^{\rm (ACC)} \big|_{B=|\Gc|=1}$ $=\frac{\rho}{\ln 2}+ o(\rho)$, whereas from Proposition \ref{prop:collapse_caching_gains} it follows that $\lim_{\rho\to 0} \frac{\Bar{R}^{\rm (MN)}}{\Bar{R}^{\rm (TDM)}} = 1$. 
			These results yield the desired $\Bar{R}^\emn = \frac{\rho}{\ln 2}   + o(\rho)$ and 
			$\lim_{\rho\to 0}\frac{\Bar{R}^{\rm (ACC)}}{\Bar{R}^{\rm (MN)}} 
			     = \lim_{\rho\to 0} \frac{\frac{\rho |\Gc|}{B \ln2}\, \Psi_{|\Gc|} + o(\rho)}{\frac{\rho}{\ln 2}   + o(\rho) }  
			     = \frac{|\Gc|}{B} \Psi_{|\Gc|}. $ \qed

    \subsection{Proof of Lemma~\ref{effectgainTDM}}\label{app:proof_lemma_effectiveGain}
      We want to prove that $\lim_{B\to\infty}\frac{\Bar{R}^{\rm (ACC)}}{\Bar{R}^{\rm (TDM)}} = \Lambda\gamma + 1$ for a fixed number of caches~$\Lambda$ and for any~$\rho$.
			Since $\E\left\{| \ln\left(1+{\rm SNR}_{g,b}\right)| \right\}<\infty$, the Strong Law of Large Numbers implies that 		
					\begin{align}\label{Limit_largeB}
					\frac{1}{B} \sum\nolimits_{b=1}^B \ln \left( 1+ {\rm SNR}_{g,b} \right) \ \stackrel{a.s.}{\longrightarrow} \ \E\left\{ \ln\left(1+{\rm SNR}_{g,b}\right) \right\}, \quad \text{as } B \to \infty,
				\end{align} 
					which implies that 
					\begin{align}\label{Limit_largeB2}
					\lim_{B\to\infty}  \frac{1}{B} \sum\nolimits_{b=1}^B \ln \left( 1+ {\rm SNR}_{g,b} \right) = \E\left\{\ln\left(1+{\rm SNR}_{g,b}\right)\right\},
				\end{align}			
			except for zero-probability events. Then since $\ln(1+x)\leq x$ $\forall x  >0$, we get that
					\begin{align}\label{Limit_largeB3}
					\E_H \Big\{ \min_{g\in\Gc}\frac{1}{B}\sum\nolimits_{b=1}^B\ln(1+\SNR_{g,b}) \Big\} \leq \E_H \Big\{ \frac{1}{B}\sum\nolimits_{b=1}^B\SNR_{g,b} \Big\} \overset{(a)}{=} \rho<\infty,
				\end{align}		
			where $(a)$ comes from the fact that $\SNR_{g,b}\sim\Exp(\frac{1}{\rho})$ and thus $\frac{1}{B}\sum_{b=1}^B\SNR_{g,b}\sim {\rm Gamma}(B,\frac{B}{\rho})$.  
			From~\eqref{Limit_largeB2} and~\eqref{Limit_largeB3}, we can apply Lebesgue's Dominated Convergence Theorem\cite[Thm.~16.4]{Billingsley1995} to interchange the order of expectation and limit and show that
			\begin{align}
					\lim_{B\to\infty} \frac{\Bar{R}^{\rm (ACC)}}{\Bar{R}^{\rm (TDM)}} 
						 & \overset{(a)}{=}  \frac{\lim_{B\to\infty}\frac{|\Gc|}{\ln 2} \E_H \left\{ \min_{g\in\Gc}\frac{1}{B}\sum_{b=1}^B\ln(1+\SNR_{g,b}) \right\}}{\frac{1}{\ln 2}\E_H\left\{ \ln \left( 1+ {\rm SNR}_{g,b} \right) \right\}} \label{Limit_largeB4a}\\
						& \overset{(b)}{=}   |\Gc|\ \frac{\E_H \left\{ \min_{g\in\Gc}\lim_{B\to\infty}\frac{1}{B}\sum_{b=1}^B\ln(1+\SNR_{g,b}) \right\}}{\E_H\left\{ \ln \left( 1+ {\rm SNR}_{g,b} \right) \right\}} 
					 \overset{(c)}{=} |\Gc| =\Lambda\gamma + 1,\label{Limit_largeB4c}
			\end{align}
			where $(a)$ follows from substituting $\bR^\eacc$ and $\bR^\emn$ by their respective expressions, 
			$(b)$ comes from the Dominated Convergence Theorem and the fact that the minimum of several continuous functions is a continuous function, and $(c)$ is due to~\eqref{Limit_largeB2}.  \qed

    \subsection{Proof of Lemma~\ref{codedgain}}\label{app:proof_lemma_effectiveGainMN}
        From~\eqref{Limit_largeB}, and by applying the same steps as in~\eqref{Limit_largeB4a}--\eqref{Limit_largeB4c}, we obtain~\eqref{eq:delta_approx_largeB_a} as
				\begin{align}\label{app_proof_last3}
					\lim_{B\to\infty}\frac{\Bar{R}^{\rm (ACC)}}{\Bar{R}^{\rm (MN)}} 
						&= \frac{\frac{|\Gc|}{\ln 2}\E_H\left\{ \ln\left(1+{\rm SNR}_{g,b}\right) \right\}} {\frac{|\Gc|}{\ln 2}\E_H\left\{ \ln \left( 1+\min_{g \in \mathcal{G}} \left\{{\rm SNR}_{g,b}\right\} \right) \right\}} 
						\mathop  =  \limits^{(a)}\exp\left(\frac{1}{\rho} -\frac{|\Gc|}{\rho}\right)
						\frac{{\rm Ei}\left(-\frac{1}{\rho}\right)}{{\rm Ei}\left(-\frac{|\Gc|}{\rho}\right)}, 
				\end{align}
			where $(a)$ follows from~\eqref{EAR_MN_Intro}. 
			To prove~\eqref{eq:delta_approx_largeB_b}, we first obtain  from~\eqref{app_proof_last3}  that
				\begin{align}
					\lim_{\rho\to 0}\lim_{B\to\infty}\frac{\Bar{R}^{\rm (ACC)}}{\Bar{R}^{\rm (MN)}} 
					& = \lim_{\rho\to 0}\exp\left(\frac{1}{\rho} -\frac{|\Gc|}{\rho}\right)
						\frac{\Ei\left(-\frac{1}{\rho}\right)}{{\rm Ei}\left(-\frac{|\Gc|}{\rho}\right)}.  \label{eq:proof_like_mn}
				\end{align}
				Then, in a similar manner as for the proof of Proposition~\ref{prop:collapse_caching_gains} in Appendix~\ref{app:gain_mn_approxLow}, we can apply the relations $-e^{-x}\ln(1+\frac{1}{x})< {\rm Ei}(-x) < \frac{-e^{-x}}{2}\ln(1+\frac{2}{x})$\cite{abramowitz55handbook} in~\eqref{eq:proof_like_mn} to obtain that 
				\begin{align}
					\lim_{\rho\to 0}\lim_{B\to\infty}\frac{\racc}{\rmn}
							& \leq  \lim_{\rho\to 0}\exp\left(\frac{1}{\rho} -\frac{|\Gc|}{\rho}\right)
						\frac{\frac{1}{2}\exp\left(\frac{-1}{\rho}\right)\ln (1+2\rho)}{\exp\left(\frac{-|\Gc|}{\rho}\right)\ln (1+\frac{\rho}{|\Gc|})} = |\Gc| \\
					\lim_{\rho\to 0}\lim_{B\to\infty}\frac{\racc}{\rmn}
							& \geq  \lim_{\rho\to 0}\exp\left(\frac{1}{\rho} -\frac{|\Gc|}{\rho}\right)
							\frac{\exp\left(\frac{-1}{\rho}\right)\ln (1+\rho)}{\frac{1}{2}\exp\left(\frac{-|\Gc|}{\rho}\right)\ln (1+\frac{2\rho}{|\Gc|})} = |\Gc|,
				\end{align}
			which concludes the proof of Lemma~\ref{codedgain}. 
			We could also obtain~\eqref{eq:delta_approx_largeB_b} by combining Proposition~\ref{prop:collapse_caching_gains} and Lemma~\ref{effectgainTDM}.	\qed

	\section{Proof of Lemma~\ref{lem:approx_expect_rate_BS}}\label{app:proof_lem_approx_expect_BS}
    To prove Lemma~\ref{lem:approx_expect_rate_BS}, we first derive the approximation in~\eqref{R_min_exact}. Afterward, we obtain the values of $\mu$ and $\sigma$ in \eqref{eq:mean_prop_bs} and \eqref{eq:var_prop_bs}, and finally we derive the integral expression of $H_{|\Gc|}$ in~\eqref{H_int}. 
    
    \subsection{Approximation for the Rate of the ACC Scheme}
    
			Let $A_{g}  \triangleq \frac{1}{B} \sum_{b=1}^B \ln \left( 1+ {\rm SNR}_{g,b} \right) =  \frac{1}{B} S_g$, for any $g\in[\Lambda]$, represent the arithmetic mean of the user capacity over the set of $B$ users of group $g$, normalized by $\ln(2)$. 
			Let us consider the Central Limit Theorem (CLT) in the large $B$ case. According to the Lindeberg-L\'evy CLT \cite{Inlow}, we have that $A_{g} \stackrel{d.}{\longrightarrow} \mathcal{N}\left(\mu, \frac{\sigma^2}{B}\right)$ as $B \to \infty,$ 
			where $d.$ stands for \emph{convergence in distribution}, and where $\mu=\E\left\{\ln \left( 1+ {\rm SNR}_{g,b} \right)\right\}$ and $\sigma^2={\rm Var}\left\{\ln \left( 1+ {\rm SNR}_{g,b} \right)\right\}$\footnote{Note that, if we focused on the low-SNR region, we could apply the approximations $\mu\approx\E\{{\rm SNR}_{g,b}\}$ and $\sigma^2\approx{\rm Var}\left\{{\rm SNR}_{g,b}\right\}$. We do not consider them here for sake of generality, and our approximation holds for any value of SNR.}. 
			We consider now the average rate for the ACC scheme when $B \to \infty$. Recall that $A_{1},  \cdots, A_{|\Gc|}$ are i.i.d. normal random variables with mean $\mu$ and variance $\nicefrac{\sigma^2}{B}$. 
			Although convergence in distribution does not generally imply convergence in mean, it was shown in~\cite{Pickands1968} that this indeed holds in the specific case of extreme values of i.i.d. random variables.  
			Consequently, $\bR^\eacc$ is given by
				\begin{align}\label{EAR_BS_LargeB}
					\lim_{B\to\infty}\Bar{R}^{\rm (ACC)} = \frac{|\Gc|}{\ln 2}\E \left\{ \min \left\{A_{1}, \cdots, A_{|\Gc|} \right\} \right\}. 
				\end{align}
			Deriving a simple closed-form expression for~\eqref{EAR_BS_LargeB} is challenging. Consequently, we propose a simple method to obtain an approximation to this expectation. 
			Since $B\to\infty$ and $A_{1},\,\cdots,\,A_{|\Gc|}$ are i.i.d. normal random variables, we can write each $A_{i}$, $i\in[|\Gc|]$, as $A_{i}=\mu+\tfrac{\sigma}{\sqrt{{B}}} A_{i}^\prime$, where $A_{i}^\prime \sim \mathcal{N}(0,1)$. Then, the minimum of $A_{1},\cdots,A_{|\Gc|}$ is re-written as
				\begin{align}\label{min_X_BS}
					&\min_{i\in|\Gc|}\left\{A_{i}\right\}  = \mu+\frac{\sigma}{\sqrt{B}} \min_{i\in|\Gc|} \left\{A_{i}^\prime\right\}.
				\end{align}
    	Then~\eqref{R_min_exact} is obtained by taking the expectation of both sides, multiplying~\eqref{min_X_BS} by  $\frac{|\Gc|}{\ln 2}$, and recalling that $H_{|\Gc|} \triangleq - \E\left\{\min_{i\in|\Gc|} \left\{A_{i}^\prime\right\}\right\}$, as defined in Section~\ref{subse:approx_largeB}. \qed
    
    \subsection{Proof of~\eqref{eq:mean_prop_bs} and \eqref{eq:var_prop_bs}  -- Mean and Variance of $\ln(1+{\rm SNR}_{g,b})$}
    
        We derive now the  expressions for $\mu$ in~\eqref{eq:mean_prop_bs} and $\sigma$ in~\eqref{eq:var_prop_bs}. 
    	Note that $\frac{\mu}{\ln(2)} \!\!=\!\!\E\left\{{\log_2 (1+{\rm SNR}_{g,b})} \right\}$ is exactly $\bar R^{\rm (TDM)}$, so we have~\eqref{eq:mean_prop_bs} by considering \eqref{EAR_MN_Intro} with $|\Gc|=1$. 
    	Moreover, we have that
    		\begin{align}\label{second_logSNR}
    			\E\left\{\left(\ln(1+{\rm SNR}_{g,b})\right)^2\right\}=\frac{1}{\rho}\int_{0}^\infty \!\left(\ln(1+x)\right)^2 \exp\left({-\frac{x}{\rho}}\right)\dd x.
    		\end{align}
    	To obtain a closed-form expression for \eqref{second_logSNR}, we re-write both the logarithmic function and the exponential function into their Meijer's G-function forms~\cite[Eq. (9.301)]{Gradshteyn}, given by
    			$\ln(1+x) = {\rm G}^{1,2}_{2,2}\left(x \left|^{1,1}_{1,0} \right.\right)$ and 
    			$\exp\left(-\frac{x}{\rho}\right) = {\rm G}^{1,0}_{0,1}\left(\frac{x}{\rho} \left|^{-}_0 \right.\right)$, 
    	respectively. Then, the integral form in \eqref{second_logSNR} becomes
    		\begin{align}\label{second_logSNR_final}
    			\E\left\{\left(\ln(1+{\rm SNR}_{g,b})\right)^2\right\}&=\frac{1}{\rho}\int\nolimits_{0}^\infty  {\rm G}^{1,2}_{2,2}\left(x \left|^{1,1}_{1,0} \right.\right)
    			{\rm G}^{1,2}_{2,2}\left(x \left|^{1,1}_{1,0} \right.\right) {\rm G}^{1,0}_{0,1}\left(\frac{x}{\rho} \left|^{-}_0 \right.\right) \dd x \notag\\
    			&\mathop  =  \limits^{(a)} 2 \exp\left(\frac{1}{\rho}\right) {\rm G}^{3,0}_{2,3}\left( \frac{1}{\rho} \left|^{1,1}_{0,0,0}\right. \right),
    		\end{align}
    	where $(a)$ follows from~\cite[Eq. (07.34.21.0081.01)]{Wolfram1} after basic simplifications. 
    	By combining the relationship
    			$\sigma^2 = \E\left\{\left(\ln(1+{\rm SNR}_{g,b})\right)^2\right\}-\left(\E\left\{{\ln (1+{\rm SNR}_{g,b})} \right\}\right)^2$ 
    	and \eqref{second_logSNR_final}, we  obtain~\eqref{eq:var_prop_bs}. \qed 
    
    \subsection{Proof of~\eqref{H_int}}
    
    	To derive the integral form of $H_{|\Gc|}$, we calculate the CDF of $V\triangleq\min \{A_{1}^\prime,\cdots,A_{|\Gc|}^\prime\}$ to be 
    		\begin{align}
    			F_{V}(y)= 1-\Pr \left\{\min \left\{A_{1}^\prime,\cdots,A_{|\Gc|}^\prime\right\}>y\right\} 
    			=1-\left(\Pr \left\{A_{1}^\prime>y\right\}\right)^{|\Gc|}\mathop  =  \limits^{(a)}1-\left(Q(y)\right)^{|\Gc|},
    		\end{align}
    	where $(a)$ holds because the CDF of the standard normal distribution is $F_{A_{i}^\prime}(x)=1-Q(x)$. 
    	The corresponding PDF is then derived by
    		\begin{align}
    			f_{V}(y) &= \frac{\partial F_{V}(y)}{\partial y}=-|\Gc| \left(Q(y)\right)^{|\Gc|-1} \frac{\partial Q(y)}{\partial y} 
    			\mathop  =  \limits^{(a)}\frac{1}{\sqrt{2 \pi}}|\Gc| \left(Q(y)\right)^{|\Gc|-1} \exp\left(-\frac{y^2}{2}\right), \notag
    		\end{align}
    	where $(a)$ follows from the integral form of the Q-function and by applying the Leibniz's Rule for differentiation under the integral sign. 
    	The value of $H_{|\Gc|}$ in \eqref{H_int} is then obtained by writing the  expectation of $V$ as an integral form by using the above PDF of $V$.  \qed

	\bibliographystyle{IEEEtran}				
	\bibliography{IEEEabrv,aBiblio}			
\end{document}